\documentclass[review]{elsarticle}
\pdfoutput=1
\usepackage[numbers]{natbib}
\usepackage{amsmath}
\usepackage{bm}
\usepackage{caption}
\usepackage{comment}
\usepackage[section]{placeins}
\usepackage{float}
\usepackage{caption}
\usepackage{subcaption}
\usepackage{booktabs}
\usepackage{pgfplots}
\pgfplotsset{width=8.5cm,height=8cm,compat=1.9}
\usepackage{tikz}
\newcommand*\circled[1]{\tikz[baseline=(char.base)]{
            \node[shape=circle,draw,inner sep=0.6pt] (char) {#1};}}
            
\usepackage{lineno,hyperref}
\hypersetup{
    colorlinks=true,
    linkcolor=blue,
    urlcolor=blue,
}
\modulolinenumbers[5]
\hypersetup{pdfauthor=author}
            
\usepackage{textcomp}
\usepackage[normalem]{ulem}      
\usepackage[colorinlistoftodos]{todonotes}
            
\journal{International Journal of Solids and Structures}

\begin{document}

\begin{frontmatter}

\title{Computational Analysis of Bubble-Structure Interactions in Near-Field Underwater Explosion}


\author[mymainaddress]{Wentao Ma}
\author[mymainaddress]{Xuning Zhao}
\author[mymainaddress]{Christine Gilbert}

\author[mymainaddress]{Kevin Wang\corref{mycorrespondingauthor}}
\cortext[mycorrespondingauthor]{Corresponding author}
\ead{kevinwgy@vt.edu}

\address[mymainaddress]{Department of Aerospace and Ocean Engineering, 
	Virginia Polytechnic Institute and State University,
	Blacksburg, Virginia 24061, USA}

\begin{abstract}

The response of underwater structures to a near-field explosion is coupled with the dynamics of the explosion bubble and the surrounding water. This multiphase fluid-structure interaction process is investigated in the paper using a two-dimensional model problem that features the yielding and collapse of a thin-walled aluminum cylinder. A recently developed computational framework that couples a finite volume compressible fluid dynamics solver with a finite element structural dynamics solver is employed. The fluid-structure and liquid-gas interfaces are tracked using embedded boundary and level set methods. The conservation of mass and momentum across the interfaces is enforced by solving one-dimensional bimaterial Riemann problems. The initial pressure inside the explosion bubble is varied by two orders of magnitude in different test cases. Three different modes of collapse are discovered, including an horizontal collapse (i.e.~with one lobe extending towards the explosive charge) that appears counterintuitive, yet has been observed in previous laboratory experiments.  Because of the transition of modes, the time it takes for the structure to reach self-contact does not decrease monotonically as the explosion magnitude increases. The fluid pressure and velocity fields, the bubble dynamics, and the transient structural deformation are visualized to elucidate the cause of each collapse mode and the mode transitions. The result suggests that, in addition to the incident shock wave, the second pressure pulse resulting from the contraction of the explosion bubble also has significant effect on the structure's collapse. The phase difference between the structural vibration and the bubble's expansion and contraction influences the structure's mode of collapse. Furthermore, the transient structural deformation has clear effect on the bubble dynamics, leading to a two-way interaction. A counter-jet that points away from the structural surface is observed. Compared to the liquid jets produced by bubbles collapsing near a rigid wall, this counter-jet is in the opposite direction.

\end{abstract}


\end{frontmatter}

\linenumbers

\section{Introduction}
%

Underwater explosions pose a significant threat to the structural integrity of marine vehicles, pipelines, and platforms. Accurate prediction of a structure's response to an underwater explosion event is crucial to ensuring safety while reducing the costs associated with overconservative design requirements. If the explosion occurs at a long distance from the structure (i.e.~far-field explosion), the load on the structure is dominated by the incident shock wave, which can be captured using one-dimensional hydrodynamics models~\cite{cole1948underwater,swisdak1978explosion,liu2008transient,mathew2018modeling}. If the explosion occurs near the structure (i.e.~near-field explosion), the problem becomes more complicated, as the bubble formed by the gaseous explosion products expands and contracts rapidly near the structure. Previous studies have demonstrated that the bubble dynamics, the dynamics of the surrounding liquid water, and the transient deformation of the structure depend on each other, leading to a fluid-structure interaction (FSI) process that involves a multiphase flow, shock waves, complex bubble geometry, large structural deformation, and nonlinear material behaviors (e.g., yielding, fracture)~\cite{Ikeda2012,gupta2016shock,leblanc2016near,guzas2019computational,MaOMAE2020,javier2021underwater}.

For thin-walled underwater structures, a primary failure mechanism is instability in the form of collapse. In the past, extensive research efforts have been devoted towards understanding the collapse of cylindrical shell structures (e.g.,~underwater pipelines, deep-sea submersibles) due to high hydrostatic pressure~\cite{kyriakides2007mechanics,ikeda2013implosion,FARHAT20132943,turner2013underwater,kishore2019underwater}. In comparison, knowledge about the response of this type of structures to a near-field explosion is still far from complete. A few recent studies suggest that the pulsation of the explosion bubble may have a substantial effect on the structure's collapse. For example, Gupta {\it et al.} conducted laboratory experiments on the collapse of aluminum 6061-T6 tubes in a confined environment, in which the tubes are subjected to the combined loading from a prescribed hydrostatic pressure and an explosion that strikes the tubes in the longitudinal direction~\cite{gupta2016shock}. Their measurement shows that the magnitude of the first bubble pulse --- that is, the increase of pressure due to the first contraction of the explosion bubble --- is lower than the incident shock wave, but not negligible. When the hydrostatic pressure is relatively low, the tube starts collapsing after the impact of this bubble pulse, which arrives much later than the incident shock wave. Later, Guzas {\it et al.} confirmed these findings using fluid-structure coupled simulations~\cite{guzas2019computational}. They also showed that the structural collapse may initiate after two to four cycles of bubble expansion and contraction, depending on the hydrostatic pressure. Ikeda conducted implosion experiments using the same type of aluminum tubes within a large pressure chamber, in which the tubes are subjected to a side-on explosion~\cite{Ikeda2012}. In these experiments, the initiation of structural collapse also occurs after the arrival of the first bubble pulse. The tubes collapse in mode 2 as expected. But surprisingly, the two lobes are aligned with the loading direction. In other words, the closest point on the cylinder to the explosive charge moves towards the charge. It is hypothesized that this counterintuitive phenomenon is related to the pulsation of the explosion bubble. Although the papers reviewed above (i.e.~\cite{gupta2016shock,guzas2019computational,Ikeda2012}) do not include details on the explosion bubble dynamics, the results presented therein indicate that the frequency of the bubble's pulsation can be similar to the first few vibration frequencies of the structure. Moreover, some other studies suggest that the dynamics of a bubble pulsating near a deformable surface can be significantly different from that of the same bubble near a rigid wall, which indicates a two-way interaction between the bubble and the structure. For example, Cao {\it et al.} investigated shock-induced bubble collapse near different types of solid and soft materials, showing that the material's acoustic impedance has an obvious effect~\cite{Cao2021}. Several previous studies (e.g.,~\cite{Gibson1968, Gibson1980, Blake1987, Duncan1991, Li2018}) have investigated bubbles pulsating near a deformable boundary, showing a liquid jet that forms in a direction away from the boundary, that is, in the opposite direction compared to the jets formed by bubbles near a rigid wall.

In this work, we investigate the response of an underwater, thin-walled aluminum cylinder to a near-field explosion, focusing on the interaction between the pulsation of the explosion bubble and the deformation and collapse of the cylinder. Clearly, the magnitude of the explosion is a key parameter. A specific objective in this study is to sweep this parameter through a broad range bounded by two extreme values: a low magnitude that does not trigger the collapse of the cylinder, and a high magnitude that causes the cylinder to collapse immediately upon the arrival of the incident shock wave. Based on the research findings mentioned above, we expect to discover transitions between different types of structural and bubble behaviors.

A recently developed fluid-structure coupled computational framework is employed in this study~\cite{Wang2015,Wang2011,Farhat2012FIVER,Main2017,HUANG2018Family}. The framework couples a finite volume compressible fluid dynamics solver with a nonlinear finite element structural dynamics solver using  a partitioned time-integration procedure~\cite{Farhat2010}. An embedded boundary method is utilized to track the wetted surface of the structure (i.e. the fluid-structure interface), which is capable of handling large structural deformation and topological changes~\cite{Wang2012ComputationalAF,Wang2015}. A level set method is utilized to track the bubble surface (i.e.~the liquid-gas interface)~\cite{FARHAT2008ghostFluid,Main2017}. The fluid-structure and liquid-gas interface conditions are enforced by the FInite Volume method with Exact two-material Riemann problems (FIVER), which naturally accommodates the propagation of shock waves across the interfaces~\cite{Wang2011,Farhat2012FIVER,Main2017}. This computational framework has been verified and validated for several multiphase flow and FSI problems that are closely related to the current application~\cite{Wang2011,FARHAT20132943,Main2017,wang2017multiphase,cao2019shock,Cao2021,xiang2021variations}. For example, Farhat {\it et al.} simulated the collapse of aluminum 6061-T6 tubes in modes 2 and 4 due to hydrostatic pressure~\cite{FARHAT20132943}. They showed that the simulation result is in close agreement with the experimental data in both the transient structural deformation and the pressure pulse generated by the structure's self-contact. Cao {\it et al.} simulated the collapse of a bubble in free field and near different solid and soft materials~\cite{Cao2021}. They showed that the simulated bubble dynamics in free field matches closely the experimental data, and the pressure time-history obtained from a bubble collapsing near a rigid wall agrees well with earlier simulations conducted using a different solver.

In this work, we consider aluminum 6061-T6 as the structural material. To properly account for geometric and material nonlinearities, the computational structural model is based on Green-Lagrange strain tensor and the $J_2$ flow theory with isotropic hardening. The elastic and plastic properties of the aluminum material are set to be the same as in the validation study presented in Farhat {\it et al.}~\cite{FARHAT20132943}. To capture the progressive yielding through the wall of the aluminum cylinder, we resolve the wall thickness explicitly in the finite element mesh. We analyze a two-dimensional model that contains one cross section of the cylinder. This geometric simplification is adopted in many studies on cylinder instabilities (e.g.,~\cite{LEBLOND201056,IAKOVLEV20081077, IAKOVLEV20081098, IAKOVLEV2009401}). In this work, it allows us to perform the aforementioned parameter sweep with mesh convergence at reasonable computational cost. The detonation process is not simulated explicitly. Instead, we initiate the simulations with a small bubble that models the state of the explosion bubble at the end of the detonation process. By varying the pressure (and hence, enthalpy) inside this bubble, we model explosions of different magnitudes. For each simulation, we visualize the fluid pressure and velocity fields, the bubble dynamics, and the transient deformation of the structure. Results from different test cases are contrasted to investigate the impact of the explosion magnitude on the dynamics of the structure and the bubble.

The remainder of this paper is organized as follows.  Section~\ref{sec:equations} presents the physical models and numerical methods employed in this work, and the setup of our simulations. Section~\ref{sec:convergence} presents a mesh convergence analysis that allowed us to determine the mesh resolution for this study. It also shows that upon convergence, the mode of collapse reported in Ikeda~\cite{Ikeda2012} is replicated. In Section~\ref{sec:examples}, we present five representative test cases with different initial pressures inside the bubble, which led to drastically different structural behaviors ranging from cyclic elastic vibration to an immediate collapse without vibration. In Section~\ref{sec:transition}, we categorize the different collapse behaviors observed in the parametric study into three modes, and discuss the cause of each mode as well as the mechanisms underlying the mode transitions. Finally, a few concluding remarks are provided in Section~\ref{sec:conclusion}.

\section{Physical models and numerical methods}
\label{sec:equations}

\subsection{Physical models}

\begin{figure}[!bht]
	\centering
		\includegraphics[width=85mm,trim={0cm 0cm 0cm 0cm},clip]{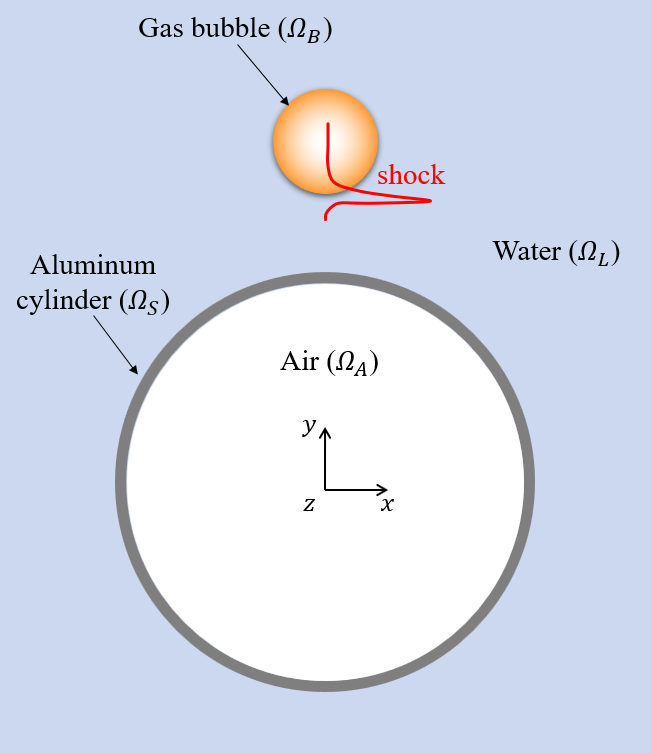}
	\caption{A two-dimensional model of an underwater aluminum cylinder subjected to a near-field explosion. (The $z$ axis is aligned with the longitudinal direction of the cylinder.)}
	\label{fig:modelProblem}
\end{figure}

Figure~\ref{fig:modelProblem} presents an illustration of the problem investigated in this work. A thin-walled, air-filled circular cylinder is submerged in water. An underwater explosion occurs in the close proximity of the cylinder, and is modeled as a gas bubble with high internal pressure and density. Let $\mathit{\Omega_S}$, $\mathit{\Omega_L}$, $\mathit{\Omega_A}$, and $\mathit{\Omega_B}$ denote the subdomains occupied by the aluminum material, the liquid water, the air inside the cylinder, and the gas bubble, respectively. The multiphase fluid flow is dominated by shock waves and high pressure. In comparison, viscous stresses and heat diffusion can be neglected. Therefore, the following Euler equations are solved in $\mathit{\Omega_L}$, $\mathit{\Omega_A}$, and $\mathit{\Omega_B}$. 
\begin{equation}
\frac{\partial \bm{W}(x,t)}{\partial t} + \nabla\cdot \mathcal{F}(\bm{W}) = \bm{0},~~~~\forall~\bm{x} \in \mathit{\Omega_L}(t)~ \cup~\mathit{\Omega_A}(t)~\cup~\mathit{\Omega_B}(t),~t>0,
\label{eq:Euler_equations}
\end{equation}
with
\begin{equation}
\bm{W}=\left[ \begin{matrix}
\rho \\
\rho \bm{V} \\
\rho e_t
\end{matrix}
\right],~~~~
\mathcal{F}=\left[ \begin{matrix}
\rho \bm{V}^T \\
\rho \bm{V} \otimes \bm{V} + p \bm{I} \\
(\rho e_t + p)\bm{V}^T
\end{matrix}
\right].
\label{eq:variables_of_Euler}
\end{equation}
Here, $\rho$, $e_t$, and $p$ denote the mass density, total energy per unit mass, and pressure, respectively.
 $\bm{V}$ is the velocity vector. $\bm{I}$ denotes the the $3 \times 3$ identity matrix.
 \begin{equation}
 e_t = e + \dfrac{1}{2}|V|^2,
 \end{equation}
 where $e$ is the internal energy per unit mass.

Equation~\eqref{eq:Euler_equations} is closed by an equation of state (EOS) for each fluid material. For the gas inside the bubble and the air inside the cylinder, we apply the perfect gas EOS, i.e.
\begin{equation}
p = (\gamma - 1) \rho e,
\label{eq:pefectGas}
\end{equation}
where $\gamma$ is the heat capacity ratio. For the liquid water, we apply the Tait EOS,  
\begin{equation}
p = p_c + \alpha \left( \left(\frac{\rho}{\rho_c}\right)^\beta - 1 \right),
\label{eq:Tait}
\end{equation}
where $\alpha = 3.5291 \times 10^8~\text{Pa}$ and $\beta = 6.4762$~\cite{Wang2015}. $\left( \rho_c, p_c \right)$ is a reference state, which is set by $\rho_c = 1.0 \times 10^3~\text{kg/m$^3$}$ and $p_c = 1.0 \times 10^5~\text{Pa}$ in this work. For each equation of state $p(\rho,e)$, the speed of sound, $c$, is given by
\begin{equation}
c = \sqrt{\left.\frac{\partial{p}}{\partial{\rho}}\right\vert_{e} + \frac{p}{\rho^2}\left.\frac{\partial{p}}{\partial{e}}\right\vert_{\rho}}.
\label{eq:soundSpeed}
\end{equation}

Within the solid subdomain, $\mathit{\Omega_S}$, the dynamic equilibrium of the cylinder undergoing finite deformation is modeled in the Lagrangian setting \cite{Cao2018_RobinNeumann}, i.e.
\begin{equation}
\rho_s \ddot{\bm{u}} \left( \bm{X}, t\right) - \nabla \cdot \left( J^{-1} \bm{F} \cdot \bm{S} \cdot \bm{F}^{T} \right) = \bm{b},~~\forall~\bm{X} \in \mathit{\Omega_S}(0),~t>0.
\label{eq:structure_equilibrium}
\end{equation}
Here, $\rho_s$ denotes the solid material's density, $\bm{u}$ the displacement vector, $\bm{X}$ the material coordinates, $\bm{S}$ the second Piola-Kirchhoff stress tensor, $\bm{F}$ the deformation gradient, and $J=\text{det}\bm{F}$. $\bm{b}$ is the body force vector acting on the cylinder, which is assumed to be zero in this study. The dots above $\bm{u}$ indicate its partial derivative with respect to time.   

The cylinder is assumed to be made of aluminum alloy 6061-T6, and can undergo yielding and plastic deformation. Following Farhat {\it et al.}~\cite{FARHAT20132943}, it is modeled as an elastic-plastic material, using the $J_2$ flow theory with isotropic hardening. The yield criterion is defined by
\begin{equation}
\sqrt{2 J_2 \left( \bm{s} \right)} = \left( \frac{3}{2} \bm{s} \cdot \bm{s} \right) ^ {1/2} = \sigma_e,
\label{eq:J2}
\end{equation}
where $\bm{s}$ is the deviator of the second Piola-Kirchhoff stress tensor, and $\sigma_e$ the von Mises effective stress.

The fluid-structure interface is given by
\begin{equation}
\mathit{\Gamma_{FS}} = \partial \mathit{\Omega_S}(t)~\cap~\left ( \partial \mathit{\Omega_L}(t) ~\cup~\partial \mathit{\Omega_A}(t)~\cup~\partial \mathit{\Omega_B}(t)  \right).
\end{equation}

Across the fluid-structure interface, the normal velocity and the surface traction are continuous, i.e.
\begin{equation}
\begin{matrix}
\left( \bm{V} - \dot{\bm{u}} \right) \cdot \bm{n} = 0, \\
-p\bm{n} = \bm{\sigma} \cdot \bm{n},
\end{matrix}~~~~~\text{on}~\mathit{\Gamma_{FS}},
\label{eq:FS_interface}
\end{equation}
where $\bm{n}$ denotes the unit normal to $\Gamma_{FS}$, $\bm{\sigma}$ the Cauchy stress tensor, which is related to the second Piola-Kirchhoff stress by $\bm{\sigma} = J^{-1} \bm{F} \cdot \bm{S} \cdot \bm{F}^{T} $.

The bubble surface (i.e.~liquid-gas interface) is given by
\begin{equation}
\mathit{\Gamma_{FF}} = \partial \mathit{\Omega_L}(t)~\cap~\partial \mathit{\Omega_B}(t).
\end{equation}
We assume that the two fluid materials across $\Gamma_{FF}$ are immiscible. In addition, surface tension is negligible compared to the hydrodynamic pressure. Therefore, normal velocity and pressure are assumed to be continuous across $\mathit{\Gamma_{FF}}$, i.e.
\begin{equation}
\begin{matrix}
\left( \lim\limits_{\bm{x}'\rightarrow\bm{x},~\bm{x}'\in\Omega_L}\bm{V}(\bm{x}')
~- \lim\limits_{\bm{x}'\rightarrow\bm{x},~\bm{x}'\in\Omega_B}\bm{V}(\bm{x}')
\right) \cdot \bm{n} = 0, \\
\lim\limits_{\bm{x}'\rightarrow\bm{x},~\bm{x}'\in\Omega_L}p(\bm{x}')
~= \lim\limits_{\bm{x}'\rightarrow\bm{x},~\bm{x}'\in\Omega_B}p(\bm{x}')\end{matrix}~~~~~\forall \bm{x}\in \Gamma_{FF}.
\label{eq:FF_interface}
\end{equation}

To track the evolution of the liquid-gas interface, we solve the level-set equation,
\begin{equation}
\frac{\partial \phi \left( \bm{x}, t \right)}{\partial t} + \bm{V} \cdot \nabla \phi = 0, ~~\forall~\bm{x} \in \mathit{\Omega_L} \cup \mathit{\Omega_A} \cup \mathit{\Omega_B} \cup \mathit{\Omega_S},
\label{eq:levelSet1}
\end{equation}
where $\phi \left( \bm{x}, t \right)$ denotes the level set function, initialized to be the signed shortest distance from $\bm{x}$ to the interface. In this way, the large deformation and topological changes (e.g. splitting and merging) of the bubble surface are  accommodated without the need of any special treatment.

\begin{figure}[!bht]
	\centering
		\includegraphics[width=85mm,trim={0cm 0cm 0cm 0cm},clip]{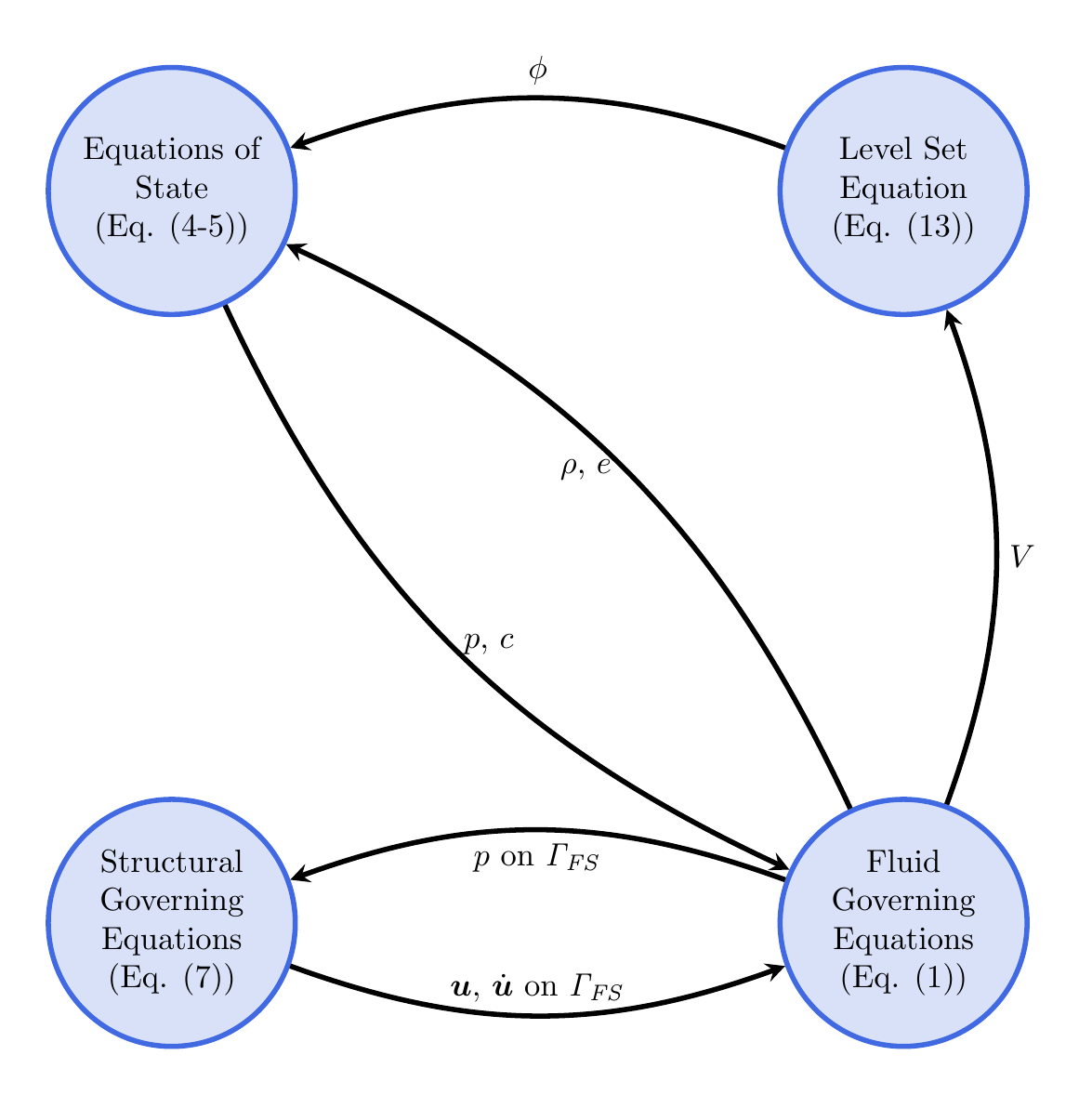}
	\caption{Physical models adopted in this work and their dependencies.}
	\label{fig:connection}
\end{figure}

In summary, Figure~\ref{fig:connection} presents an overview of the model equations solved in this work, as well as their dependencies.

\subsection{Numerical methods}
\label{sec:FSI}

A recently developed multiphase fluid-structure coupled computational framework is applied to solve the aforementioned governing equations \cite{Farhat2012FIVER, Wang2011, Wang2012ComputationalAF, Main2017}. This framework couples a nonlinear finite element solid dynamics solver with a finite volume fluid dynamics solver using a partitioned procedure.

\begin{figure*}[htb!]
\centering
\hspace*{-2.5cm}  
\includegraphics[width=170mm,trim={0cm 0cm 0cm 0cm},clip]{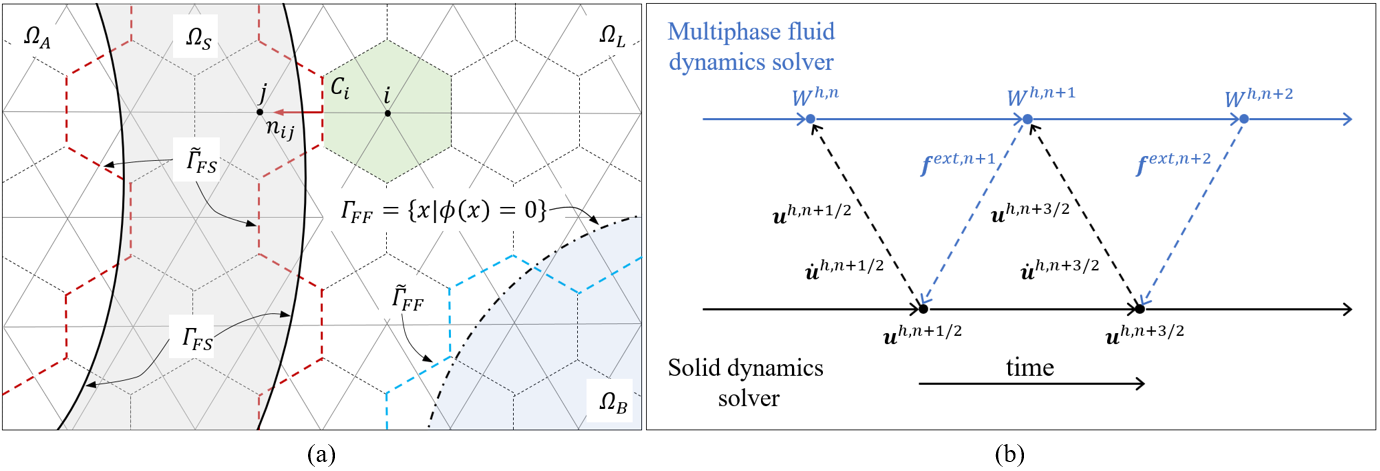}
\caption{Illustration of the discretization methods in space (a) and time (b).}
\label{fig:discretization}
\end{figure*}   

As shown in Figure~\ref{fig:discretization}(a), an augmented fluid domain $\mathit{\tilde{\Omega}}$ is defined to include the space occupied by the liquid, the gas bubble, the cylinder, and the air inside, i.e.
\begin{equation}
\mathit{\tilde{\Omega}} = \mathit{\Omega_L} \cup \mathit{\Omega_B} \cup \mathit{\Omega_S} \cup \mathit{\Omega_A}.
\end{equation}
In $\mathit{\tilde{\Omega}}$, a node-centered, unstructured, and non-interface-conforming finite volume mesh is used to semi-discretize the fluid governing equations. Around each node (e.g.~Node $i$ in Figure~\ref{fig:discretization}(a)), a control volume $C_i$ is constructed. Integrating Eq.~\eqref{eq:Euler_equations} within $C_i$ gives the semi-discrete form, 
\begin{equation}
\frac{\partial \bm{W}_i}{\partial t} + \frac{1}{\left \| C_i \right \|}\sum_{j \in N(i)}\int_{\partial C_{ij}} \mathcal{F}(\bm{W}) \cdot \bm{n_{ij}} dS =0,
\label{eq:integrating_Euler_equation}
\end{equation}
where $\bm{W}_i$ denotes the average of $\bm{W}$ in $C_i$. $\left \| C_i \right \|$ denotes the volume of $C_i$. $N(i)$ is the set of neighboring nodes that are connected to node $i$ by an edge. $\partial C_{ij} = \partial C_i~\cap~\partial C_j$ is the interface between $C_i$ and $C_j$. $\bm{n}_{ij}$ is the unit vector normal to $\partial C_{ij}$. We compute the surface integral over $\partial C_{ij}$ in different ways, depending on the location of nodes $i$ and $j$ --- specifically, which fluid or solid subdomain they belong to. The following four scenarios are considered.
\begin{enumerate}
\item[(1)] Nodes $i$ and $j$ are both located in the same fluid subdomain ($\mathit{\Omega_L}$ or $\mathit{\Omega_A}$ or $\mathit{\Omega_B}$). In this case, the method of monotonic upwind scheme conservation law (MUSCL)~\cite{VanLeer1979} and Roe's flux~\cite{Roe1981} are used to calculate the flux $\mathcal{F} \left( \bm{W} \right)$ across $\partial C_{ij}$.  

\item[(2)] Nodes $i$ and $j$ belong to different fluid subdomains. In this scenario, a one-dimensional (1-D) two-fluid Riemann problem is constructed along the edge $i$-$j$, that is,
\begin{equation}
\frac{\partial \bm{w}}{\partial \tau} + \frac{\partial \mathcal{F} \left( \bm{w} \right)}{\partial \xi} = 0, ~~\text{with}~ \bm{w} (\xi, 0) = 
\begin{cases}
\bm{w}_i & \text{if}~\xi \leq 0, \\
\bm{w}_j & \text{if}~\xi > 0,
\end{cases}
\label{eq:FF_Riemann}
\end{equation}
where $\tau$ denotes the time coordinate, and $\xi$ the local spatial coordinate aligned with $\bm{n}_{ij}$ and centered at the midpoint between $i$ and $j$. The initial states $\bm{w}_i$ and $\bm{w}_j$ are projections of $\bm{W}_i$ and $\bm{W}_j$ on the $\xi$ axis. This 1-D Riemann problem is solved exactly. Its solution is supplied to Roe's flux function to calculate the flux across $\partial C_{ij}$, thereby enforcing the interface conditions (Eq.~\eqref{eq:FF_interface})~\cite{Farhat2012FIVER}. 
 
\item[(3)] One of the two nodes belongs to a fluid subdomain, while the other node belongs to the solid subdomain, $\Omega_S$. In this case, a 1-D fluid-structure Riemann problem with a moving wall boundary is constructed. For example, if node $i$ is the one in a fluid subdomain, the Riemann problem is 
\begin{equation}
\frac{\partial \bm{w}}{\partial \tau} + \frac{\partial \mathcal{F} \left( \bm{w} \right)}{\partial \xi} = 0, ~~\tau > 0,~\xi < v_S \tau,
\label{eq:FS_Riemann1}
\end{equation}
\vspace{-0.7 cm}
\begin{equation}
\bm{w} \left( \xi, 0 \right) = \bm{w}_i, ~~\xi < 0,
\label{eq:FS_Riemann2}
\end{equation}
\vspace{-0.7 cm}
\begin{equation}
v \left( v_S \tau, \tau \right) = v_S, ~~\tau > 0,
\label{eq:FS_Riemann3}
\end{equation}
where $\xi$ is the local spatial coordinate along $\bm{n}_{ij}$, centered at the midpoint between $i$ and $j$. The initial state $\bm{w}_i$ is reconstructed by the fluid state $\bm{W}_i$. $v_S$ denotes the normal velocity of the structure at its intersection with edge $i-j$, which is computed by the structural dynamics solver. Similar to the previous scenario, the exact solution of this 1-D Riemann problem is supplied to Roe's flux function to calculate the flux across $\partial C_{ij}$ \cite{Wang2011, Wang2012ComputationalAF}. 

\item[(4)] Both nodes $i$ and $j$ belong to the solid subdomain. In this case, the flux across $\partial C_{ij}$ is set to $0$.
\end{enumerate}

The algorithm above is referred to as FIVER, which stands for FInite Volume method with Exact two-material Riemann problems~\cite{Wang2015, Farhat2012FIVER, Wang2011, Wang2012ComputationalAF, Main2017, FARHAT2008ghostFluid, MAIN2014implicit, HUANG2018Family}. It has been employed in the past to simulate the collapse of underwater structures due to hydrostatic pressure \cite{FARHAT20132943, WangOMAE2014, MaOMAE2020}, as well as bubble dynamics in free field and near different material boundaries \cite{Wang2017, Cao2021}. FIVER requires tracking the fluid-structure and liquid-gas interfaces in the non-interface-conforming, unstructured mesh $\mathit{\tilde{\Omega}^h}$. A collision-based computational geometry algorithm \cite{Wang2015, Wang2012ComputationalAF} is applied to track the fluid-solid interface. 

The liquid-gas interface is tracked implicitly by solving the level set equation~\eqref{eq:levelSet1}. In this study, \eqref{eq:levelSet1} is first rearranged to obtain
\begin{equation}
\frac{\partial \phi \left( \bm{x}, t \right)}{\partial t} + \nabla \cdot \left( \phi \bm{V} \right) = \phi \nabla \cdot \bm{V}.
\label{eq:levelSet2}
\end{equation} 
Equation~(\ref{eq:levelSet2}) is solved using a finite volume method on the same fluid mesh. Specifically, the convection term, $\nabla \cdot \left( \phi \bm{V} \right)$, is discretized using the same MUSCL scheme, but without a slope limiter. The term $\phi \nabla \cdot \bm{V}$ on the right-hand-side is treated as a source term. Additional details can be found in \cite{Main2017}.

A Galerkin finite element method is applied to semi-discretize the weak form of Equation~(\ref{eq:structure_equilibrium}), which yields 
\begin{equation}
\bm{M} \frac{\partial^2 \bm{u}^h}{\partial t^2} + \bm{f}^{int} \left( \bm{u}^h, \frac{\partial \bm{u}^h}{\partial t} \right) = \bm{f}^{ext},
\label{eq:Galerkin}
\end{equation}
where $\bm{M}$ denotes the mass matrix, $\bm{u}^h$ denotes the discrete displacement vector; $\bm{f}^{int}$ and $\bm{f}^{ext}$ denote the discrete internal force and external force vector, respectively.

The staggered fluid-structure time integrator presented in \cite{Farhat2010} is used in this work to integrate the coupled fluid and structural governing equations. The fluid equations are integrated in time using an explicit fourth-order Runge–Kutta method, while the structural equations are integrated using the second-order central difference scheme. Notably, the fluid and solid time steps
are offset by half a step (Figure~\ref{fig:discretization}(b)). This is a designed feature to achieve second-order accuracy in time while maintaining optimal numerical stability.

\subsection{Simulation setup}

\begin{figure}
    \centering
    \includegraphics[width=85mm,trim={0cm 0cm 0cm 0cm},clip]{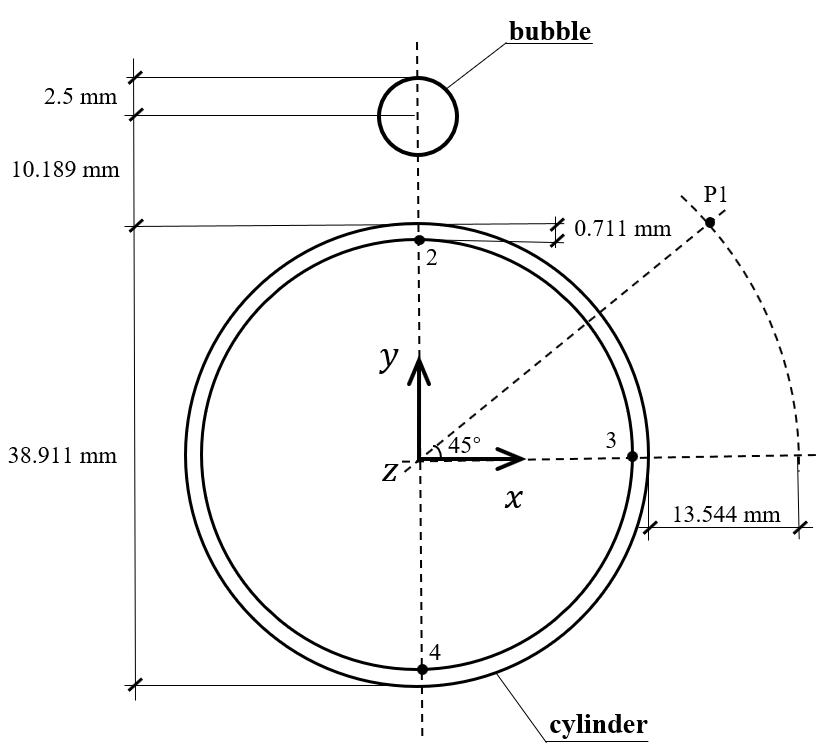}
    \caption{Setup of numerical experiment.}
    \label{fig:setup}
\end{figure}

The setup of the numerical simulations is shown in Figure~\ref{fig:setup}, including the dimensions of the cylinder and the bubble. A fluid pressure sensor, P1, is placed in the subdomain of liquid water at a location that is close to both the cylinder and the bubble. Also, three displacement and strain sensors are placed on the inner wall of the cylinder. The material and geometric properties of the cylinder are listed in Table~\ref{tab:solid}. In particular, the aluminum material properties are set to be the same as in~\cite{FARHAT20132943}. The properties of the bubble are listed in Table~\ref{tab:bubble}. Here, the stand-off distance is defined as the shortest distance between the bubble's center and the cylinder's outer surface. The bubble's initial pressure in this study is varied from $1.0~\text{MPa}$ to $100.0~\text{MPa}$ in different test cases. The properties of the ambient water and the air inside the cylinder are listed in Table~\ref{tab:fluid}. 

\begin{table}[H]
\caption{Material and geometric properties of the cylinder (Aluminum 6061-T6)}\label{tab:solid}
\hspace{-2.8cm}
\begin{tabular}{@{} ccccccc@{} }
\toprule
Young's modulus & Poisson's ratio & Density & Yield stress & Tangent modulus & Outer diameter & Thickness \\
\midrule
$69.6$ GPa & $0.33$ & $2779$ kg/$\text{m}^3$ & $292$ MPa & $674$ MPa & $38.911$ mm & $0.711$ mm \\
\bottomrule
\end{tabular}
\end{table}

\begin{table}[H]
\caption{Bubble properties}\label{tab:bubble}
\hspace{-1.3cm}
\begin{tabular}{@{} ccccc @{} }
\toprule
Stand-off distance & Initial radius & Initial density  & Initial pressure & Heat capacity ratio\\
\midrule
$10.189$ mm & $2.5$ mm & $50.0$ kg/m$^3$ & $1.0$ to $100.0$ MPa & 1.4\\
\bottomrule
\end{tabular}
\end{table}

\begin{table}[H]
\caption{Properties of the ambient water and the air inside the cylinder}\label{tab:fluid}
\hspace{-1cm}
\begin{tabular}{@{} ccccc @{} }
\toprule
Water pressure & Water density & Air pressure& Air density & Air heat capacity ratio\\
\midrule
$1.0$ MPa & $1000.39$ kg/m$^3$ & $0.1$ MPa & $1.225$ kg/m$^3$ & 1.4\\
\bottomrule
\end{tabular}
\end{table}

\section{Mesh convergence analysis}
\label{sec:convergence}
\label{subsec:meshConvergence}
\begin{figure*}[htb!]
\centering
\hspace*{-2.5cm}  
\includegraphics[width=170mm,trim={0cm 0cm 0cm 0cm},clip]{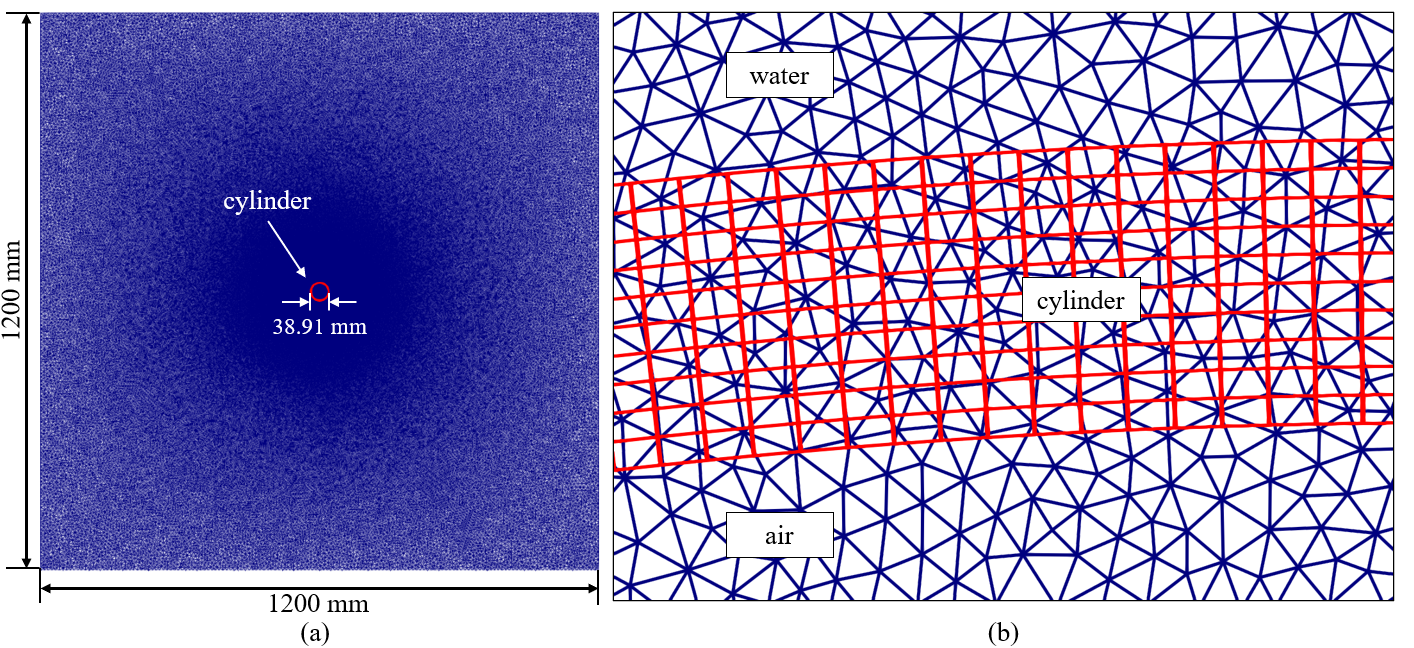}
\caption{Example fluid and structural meshes (Pair 8): the entire computational domain (a) and a zoom-in snapshot around the fluid-structure interface (b).}
\label{fig:sampleMesh}
\end{figure*}

A test case with initial pressure, $p_0 = 12.5$ MPa, inside the bubble is selected as an example problem to demonstrate the capability of achieving mesh convergence and to find appropriate mesh resolutions for the parametric study presented in Sections~\ref{sec:examples} and~\ref{sec:transition}. Eight ($8$) pairs of fluid and structural meshes were created for this test case, with resolution varying by a factor of $10$. All the fluid meshes are unstructured and nonuniform, mostly refined in a circular region that contains both the structure and the bubble (at its maximum size). Table~\ref{tab:grids} summarizes the important parameters of these meshes. As an example, Figure~\ref{fig:sampleMesh} shows the fluid and structural meshes in Pair 8. The fluid domain is a square with a length of $1,200$ mm, which is approximately $30$ times the diameter of the structure. Figure~\ref{fig:sampleMesh}(b) highlights the fact that the fluid mesh does not conform to the boundary of the structure. The embedded boundary method described in Section \ref{sec:FSI} is employed to track the structure within this fluid mesh. 

All the computations are performed using the Tinkercliffs computer cluster at Virginia Tech. The fluid dynamics solver is parallelized using Message Passing Interface (MPI). As an example, for mesh pair 8, the fluid mesh is divided into $2,047$ subdomains, each one assigned to an AMD EPYC 7702 processor core. The time step size is $3.5\times 10^{-6}~\text{ms}$. To advance the physical time by $1.0~\text{ms}$, $28.3$ hours of wall-clock time are needed, which means a computational cost of $5.8\times 10^4$ core-hours. The total computational cost of the simulation on mesh pair 8 is $2.3\times 10^5$ core-hours.

\begin{table*}[!htb]
    \centering
    \caption{Fluid and structural meshes used in the mesh convergence analysis.}
    \label{tab:grids}
    \begin{tabular}{@{}cccc@{}}
    \toprule
         & Structural Mesh & \multicolumn{2}{c}{Fluid Mesh}\\ \midrule
         & Resolution$^*$ & Num. of Nodes & Element Size$^{**}$ (mm) \\ \midrule
    Pair 1 & 100 $\times$ 1 & $4.29 \times 10^4$ & 1.2 \\ \midrule
    Pair 2 & 200 $\times$ 3 & $1.71 \times 10^5$ & 0.6 \\ \midrule
    Pair 3 & 260 $\times$ 4 & $2.99 \times 10^5$ & 0.45 \\ \midrule
    Pair 4 & 336 $\times$ 4 & $4.66 \times 10^5$ & 0.36 \\ \midrule
    Pair 5 & 400 $\times$ 5 & $6.74 \times 10^5$ & 0.3 \\ \midrule
    Pair 6 & 600 $\times$ 5 & $6.74 \times 10^5$ & 0.3 \\ \midrule
    Pair 7 & 600 $\times$ 5 & $1.51 \times 10^6$ & 0.2 \\ \midrule
    Pair 8 & 1000 $\times$ 10 & $4.19 \times 10^6$  & 0.12 \\ \bottomrule
    \end{tabular}
    \begin{tabular}{ll}
    {\footnotesize $*$} & {\footnotesize The first/second number is the number of elements to resolve the circumference/thickness} \\[-1ex]
     & {\footnotesize of the cylinder.}\\
    {\footnotesize $**$} & {\footnotesize In the most refined region.}
    \end{tabular}
\end{table*}

\begin{figure*}[!htb]
\centering
\hspace*{-0.5cm}  
    \includegraphics[width=130mm,trim={0cm 0cm 0cm 0cm},clip]{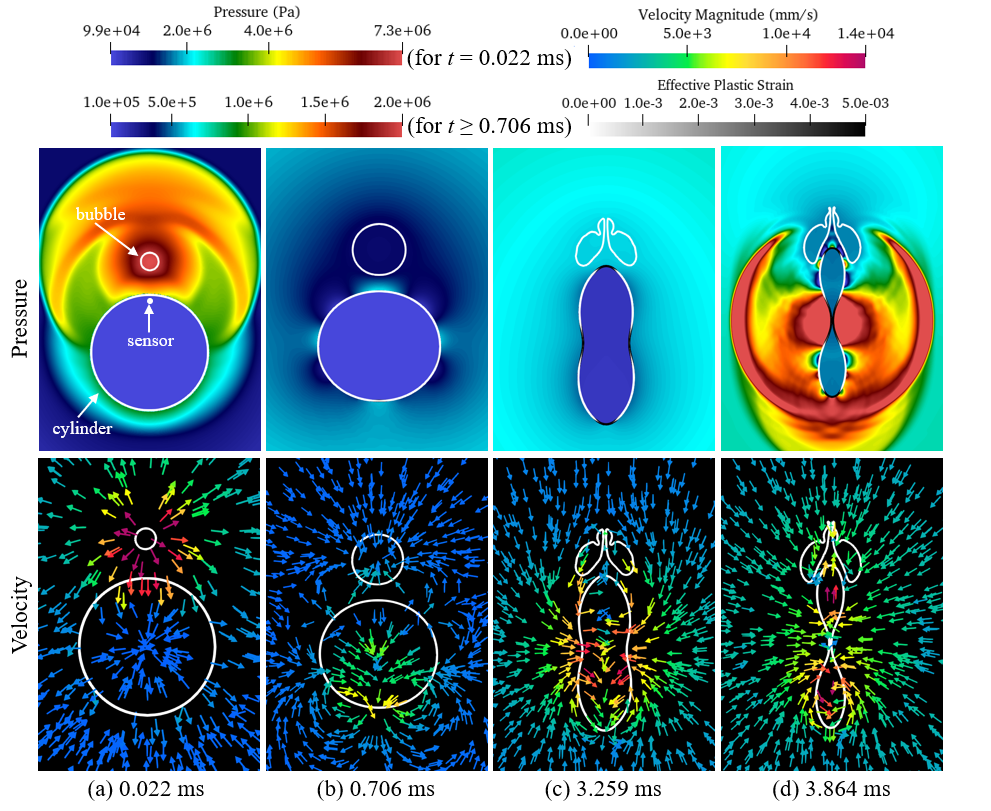}
    \caption{Results obtained using mesh pair 8 at four different time instants: The fluid pressure field and the plastic deformation of the structure (top row), and the fluid velocity field (bottom row).}
    \label{fig:flowFieldMeshConverge}
\end{figure*}

\begin{figure*}[!htb]
\centering
\hspace*{-0.5cm}  
    \includegraphics[width=110mm,trim={0cm 0cm 0cm 0cm},clip]{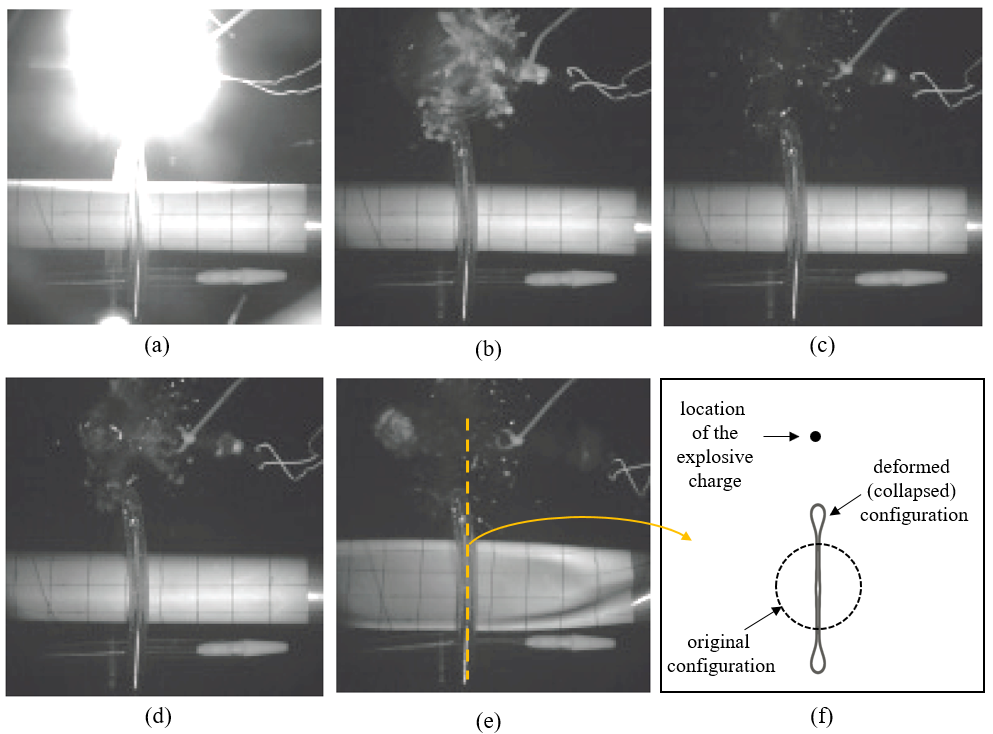}
    \caption{Experimental result of the collapse of an aluminum 6061 tube due to a near-field explosion (Ikeda~\cite{Ikeda2012}, test AE05r01). (a)-(e) A sequence of images from the high-speed movie obtained from this test. (f) Schematic drawing of the collapsed cylinder (a cross-sectional view).}
    \label{fig:expPhoto}
\end{figure*}

Figure~\ref{fig:flowFieldMeshConverge} presents four solution snapshots obtained using the finest meshes (i.e.~Pair 8). As soon as the simulation begins, the bubble generates a strong outgoing shock wave because of the high internal pressure. The first snapshot ($t = 0.022$ ms) captures the impact of this shock wave on the cylinder as well as the reflection. At the same time, the bubble starts to expand, which can be observed from the fluid velocity field. The second snapshot ($t = 0.706~\text{ms}$) is taken at a time shortly after the bubble reaches its maximum size. The third snapshot ($t = 3.259~\text{ms}$) is taken after the bubble has gone through two cycles of oscillation (i.e. expansion and contraction). A liquid jet, which points away from the structure, penetrates the bubble's top surface. At this time instant, the structure has lost stability and undergone large deformation. Consequently, plastic strain occurs on the structure. It can be observed in the top image of Figure~\ref{fig:flowFieldMeshConverge}(c) that plastic deformation is concentrated at the left, right, top, and bottom portions of the cylinder. The last snapshot is taken at $t= 3.864$ ms, shortly after the structure has reached self-contact. The emission of a shock wave at the point of contact can be clearly observed. This type of {\it implosion} shock waves have been observed and investigated in the past in the context of hydrostatic collapse~\cite{FARHAT20132943, GUPTA2014, MUTTAQIE2020, SALAZAR2020}. Notably, in the final configuration, the two lobes of of the structure extends in the vertical direction, that is, in the propagation direction of the incident shock wave. The same behavior has been observed in a previous laboratory experiment reported in Ikeda~\cite{Ikeda2012}(Figure~\ref{fig:expPhoto}). The cause of this mode of collapse will be discussed in Section~\ref{sec:examples} of this paper. 

\begin{figure}[!htb]
    \centering
    \includegraphics[width=85mm,trim={0cm 0cm 0cm 0cm},clip]{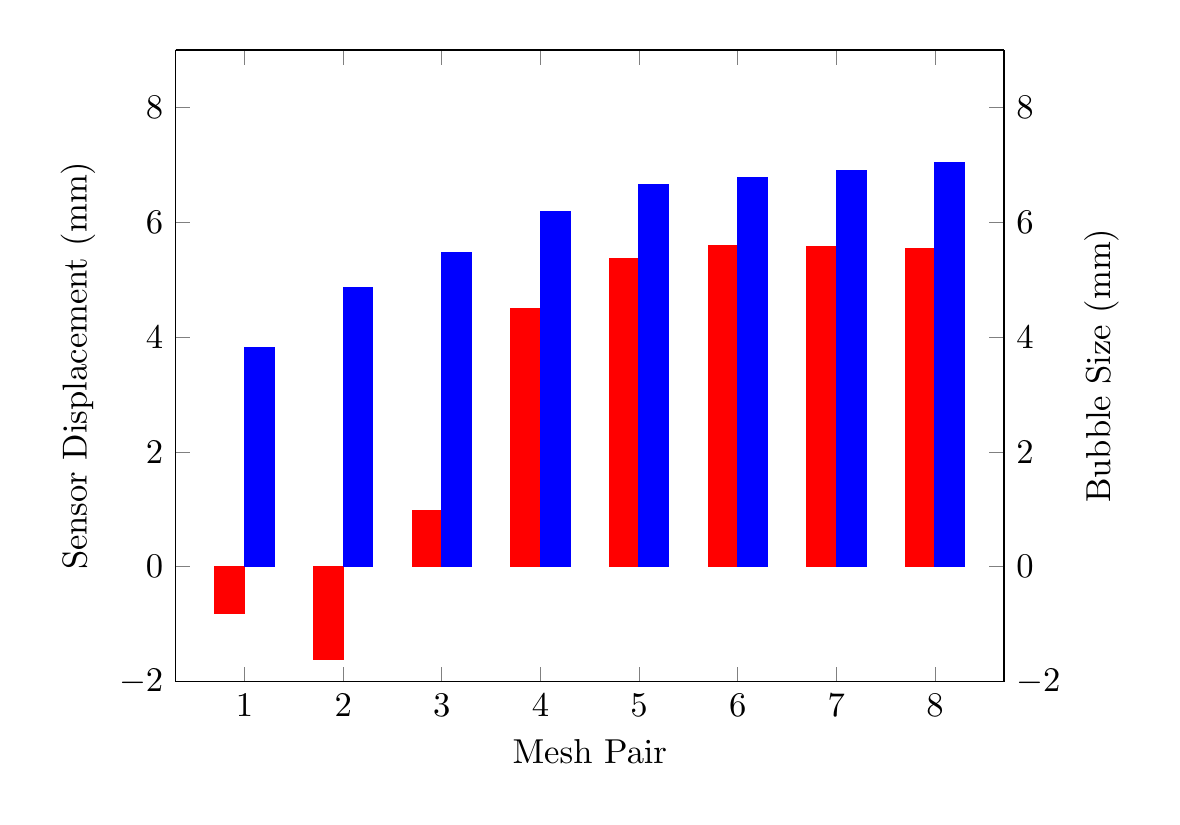}
    \caption{Mesh convergence analysis: Vertical displacement at the sensor marked in Figure~\ref{fig:flowFieldMeshConverge} (red) and bubble size (radius equivalent, blue) at 3.259 ms, obtained using different pairs of meshes. (Pair 1: coarsest, Pair 8: finest)}
    \label{fig:disp_radius_meshConverge}
\end{figure}

Figure~\ref{fig:disp_radius_meshConverge} shows the convergence of the numerical results as the mesh gets refined. Two quantities of interest are examined, namely the structural displacement at the sensor location marked in Figure~\ref{fig:flowFieldMeshConverge}, and the bubble size. Both of them are measured at $t = 3.259~\text{ms}$. As the mesh gets refined, the convergence of these quantities are achieved. From mesh pairs 4 to 8, although the computational cost is increased by $30$ times (in terms of core-hours), the result only changes $18.8\%$ in sensor displacement and $12.2\%$ in bubble size.
Based on this analysis, mesh pair 6 is selected for the parametric study in the subsequent sections of the paper. The discrepancy between the solution obtained using mesh pair 6 and that using mesh pair 8 is only $0.76\%$ in sensor displacement and $3.71\%$ in bubble size. 

\section{Bubble-structure interaction and different collapse modes}
\label{sec:examples}

To elucidate the dynamic bubble-fluid-structure interaction and the impact of this interaction on the structure's collapse, a parametric study was conducted with initial pressure inside the bubble (denoted by $p_0$) varied from $1$ MPa to $100$ MPa in different test cases, while all other parameters remained fixed. It was observed that the dynamics of the bubble and the structure do not evolve monotonically with respect to the variation of $p_0$.  In this section, we present five ($5$) representative cases that exhibit dramatically different modes, whereas the transition among these modes is discussed in Section~\ref{sec:transition}.


\subsection{$p_0 = 8.0$~MPa (enthalpy: $0.5498~\text{J/mm}$)}
\begin{figure}
\hspace*{-2.4cm}  
 \includegraphics[width=170mm,trim={0cm 0cm 0cm 0cm},clip]{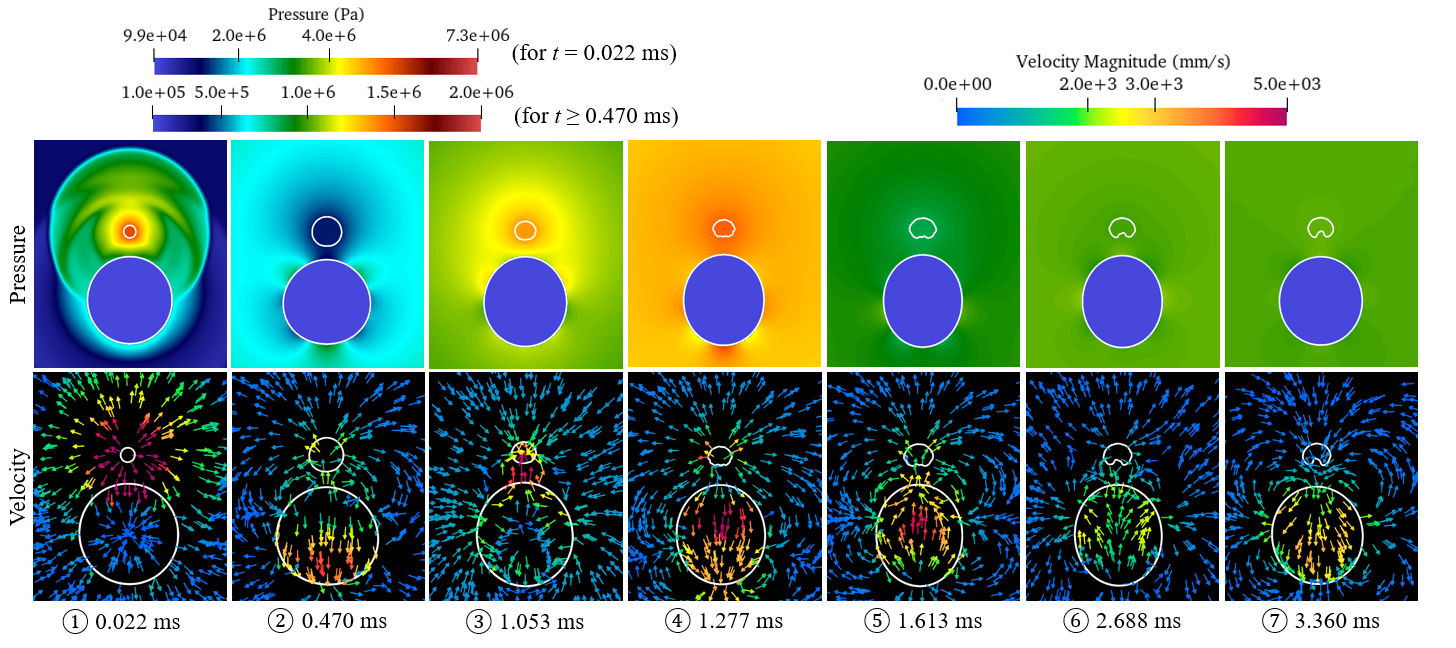}
    \caption{Snapshots of the fluid and structural results in the case of $p_0 = 8.0~\text{MPa}$.}
    \label{fig:8MPa_flowField}
\end{figure}

\begin{figure}
\begin{subfigure}{.95\textwidth}
\centering
 \includegraphics[width=85mm,trim={0cm 0cm 0cm 0cm},clip]{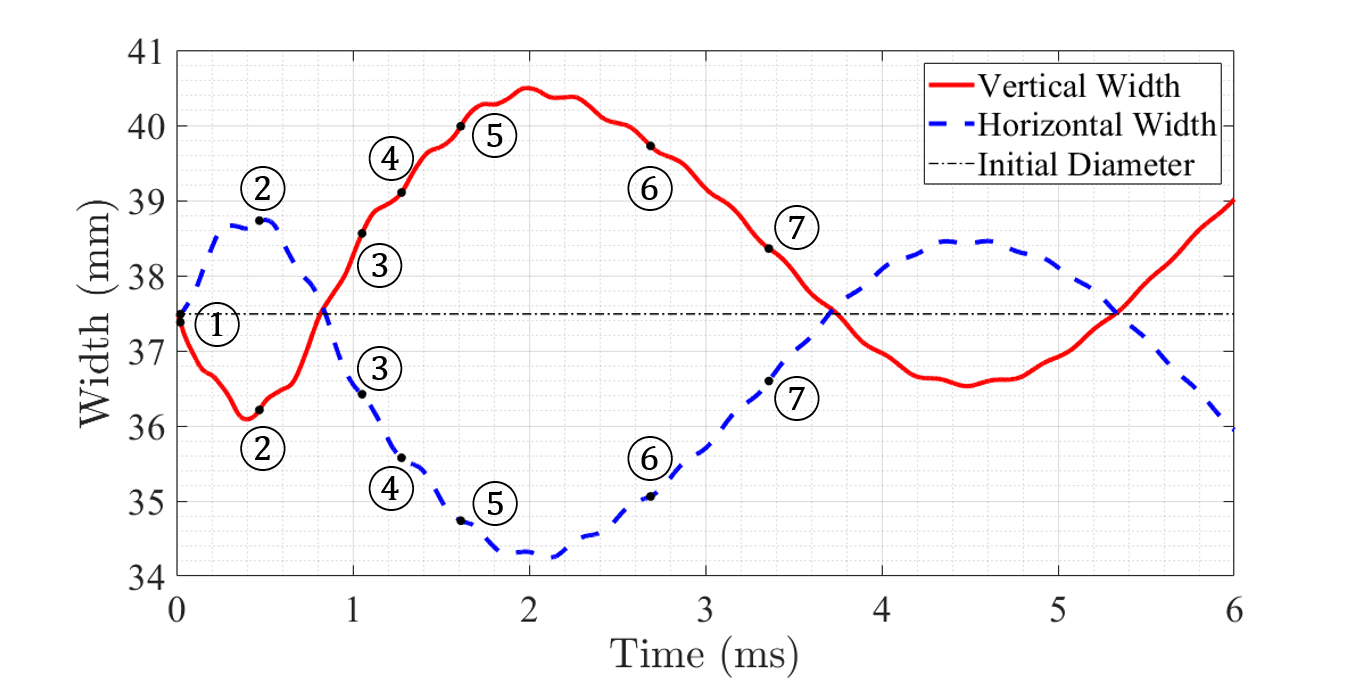}
 \caption{}
\end{subfigure}
\begin{subfigure}{.95\textwidth}
\centering
 \includegraphics[width=85mm,trim={0cm 0cm 0cm 0cm},clip]{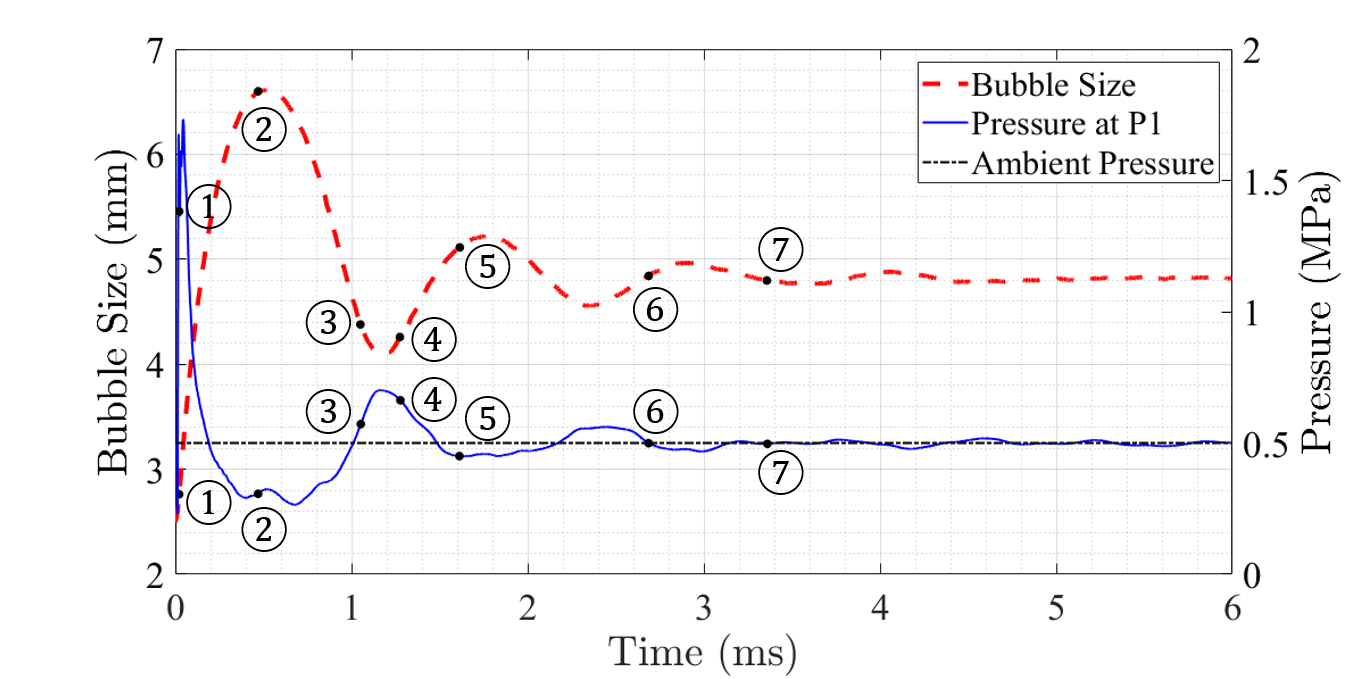}
 \caption{}
 \end{subfigure}
 \caption{Time history of selected quantities of interest in the case of $p_0 = 8.0~\text{MPa}$. (a) The distances between the top and bottom points (i.e.~vertical width)and the left and right points (i.e.~horizontal width) of the cylinder. (b) The bubble size (radius equivalent) and the fluid pressure at a sensor location. The time instants shown in Figure~\ref{fig:8MPa_flowField} are marked on these curves.}
    \label{fig:8MPa_History}
\end{figure}

In this case, the pressure load created by the bubble is not high enough to make the cylinder collapse. The dynamic process features the cyclic expansion and contraction of the bubble, coupled with the oscillation of the cylinder. Figure~\ref{fig:8MPa_flowField} presents a series of solution snapshots that show the evolution of the bubble, the cylinder, and the fluid pressure and velocity fields. Furthermore, the structural deformation is characterized by the distance between the cylinder's top and bottom points and the distance between its left and right points, as shown in Figure~\ref{fig:8MPa_History}(a). The time histories of bubble size and fluid pressure at a sensor location (P1 in Figure~\ref{fig:setup}) are shown in Figure~\ref{fig:8MPa_History}(b).

In Figure~\ref{fig:8MPa_flowField}, the first snapshot (sub-figure \circled{1}) is taken at $t = 0.022$ ms, shortly after the incident shock wave generated by the bubble reaches the surface of the cylinder. The bubble is impacted by the reflection of the incident shock wave against the cylinder, which is the first evidence of a two-way coupling between bubble and structural dynamics. The expanding bubble pushes the surrounding water. Therefore, the top portion of the cylinder is impacted by both a pressure load from the incident shock wave and a momentum from the water flow. As a result, the cylinder is compressed vertically. Around $t = 0.470~\text{ms}$ (Figure~\ref{fig:8MPa_flowField}~\circled{2}), the bubble is about to reach its maximum size. As the speed of its expansion decreases, the pressure of the surrounding water also decreases. At this time, the cylinder starts to bounce back from the vertical compression, which drives the volume of water above it to move towards the bubble. This again indicates that the bubble dynamics is affected by the transient structural deformation. After $t = 0.528~\text{ms}$, the bubble starts to contract. In accordance with the vertical stretch of the cylinder, a high pressure region occurs between the bubble and the structure at around 0.65 ms, which causes an increased vertical pressure gradient that accelerates the upward water flow at the bubble's bottom surface. Although this vertical pressure gradient gradually decreases and reverses direction at around 1 ms, the accelerated contraction of the bubble's bottom surface continues due to the inertia of water. Sub-figure~\circled{3} is taken at $t = 1.053$ ms, when the bubble is about to reach its minimum size. From the velocity field, the faster contraction of the bubble's bottom surface can still be observed. Sub-figure~\circled{3} also shows that as the bubble's contraction slows down, the local pressure increases. This increase is captured by sensor P1 as a pressure pulse around $1.2~\text{ms}$ (Figure~\ref{fig:8MPa_History}(b)). At $t = 1.193~\text{ms}$, the bubble contracts to its first minimum size and begins to expand again. Sub-figure~\circled{4} is taken at $t = 1.277~\text{ms}$. From this time onward, the bubble's shape becomes clearly non-spherical. The last three sub-figures are taken during the second and third cycles of bubble oscillation (i.e.~expansion and contraction). The pressure variation becomes smaller both in time and in space. A dent gradually develops at the bottom of the bubble, which can be attributed to both the reflection of the incident shock wave against the structural surface and the accelerated upward water flow generated by the high vertical pressure gradient between the bubble and the cylinder.

In this test case, the structural deformation is relatively small. Yielding only occurs on the outer and inner surfaces of the top, bottom, left, and right points of the cylinder, with the maximum value of effective plastic strain less than $2.0\times 10^{-3}$. Figure~\ref{fig:8MPa_History}(a) shows that although some higher frequency vibration modes are activated by the non-uniform hydrodynamic loads, the structural deformation is dominated by the first asymmetric breathing mode, which features alternative compression and expansion in vertical and horizontal directions~\cite{KWON1993,Blevins1979}. This figure also indicates that both the mean configuration of the cylinder and the configuration with the largest deformation (at around $2.0~\text{ms}$) have an ellipsoidal shape, with the primary (longer) axis in the vertical direction. In other words, the result indicates a tendency of collapsing into a horizontally compressed configuration.

\subsection{$p_0 = 12.5$~MPa (enthalpy $0.8590~\text{J/mm}$)}

\begin{figure*}[!htb]
\hspace*{-2.5cm}  
 \includegraphics[width=170mm,trim={0cm 0cm 0cm 0cm},clip]{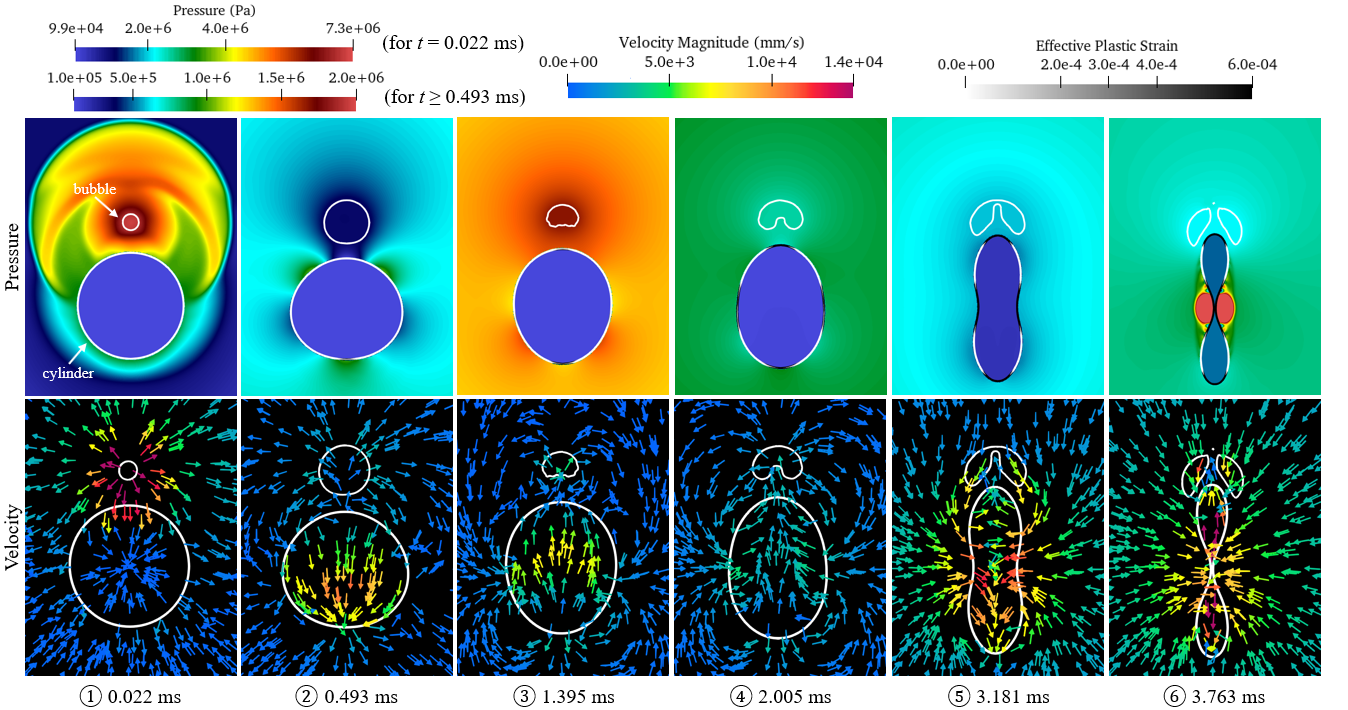}
    \caption{Snapshots of the fluid and structural results in the case of $p_0 = 12.5~\text{MPa}$.}
    \label{fig:12_5MPa_flowField}
\end{figure*}

\begin{figure}[!htb]
\begin{subfigure}{.95\textwidth}
\centering
 \includegraphics[width=85mm,trim={0cm 0cm 0cm 0cm},clip]{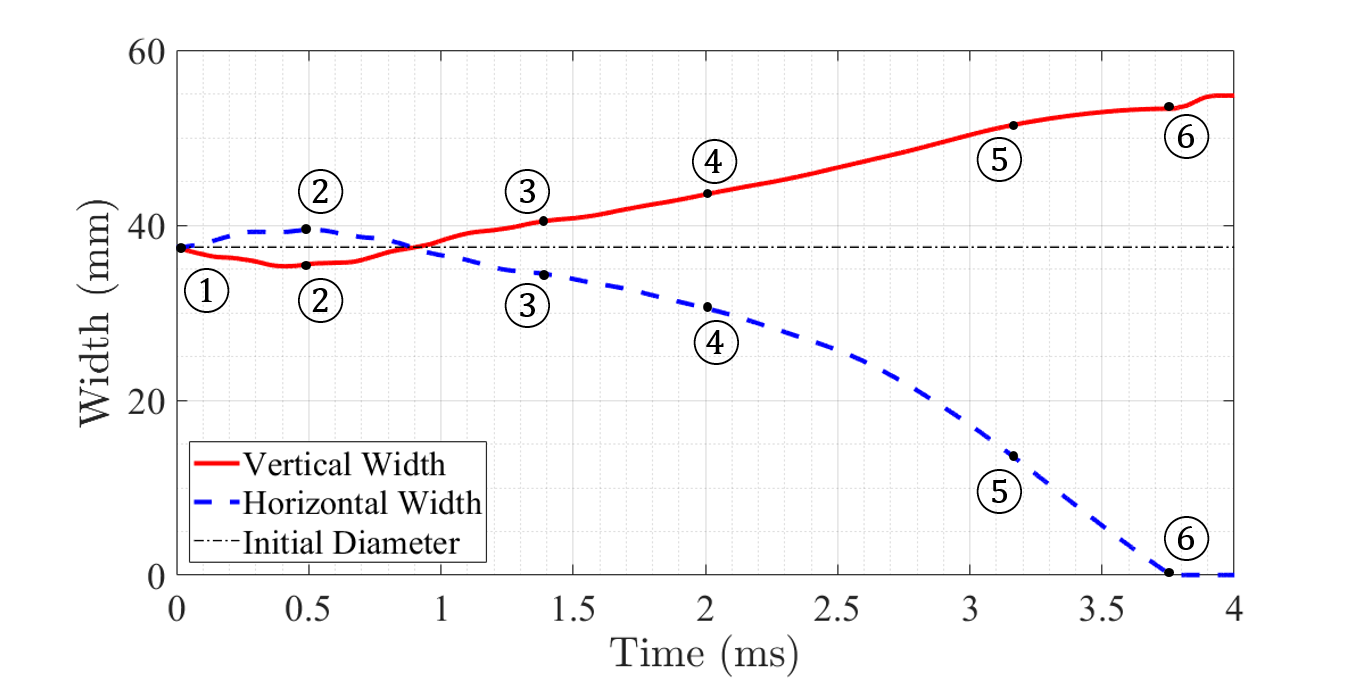}
 \caption{}
\end{subfigure}
\begin{subfigure}{.95\textwidth}
\centering
 \includegraphics[width=85mm,trim={0cm 0cm 0cm 0cm},clip]{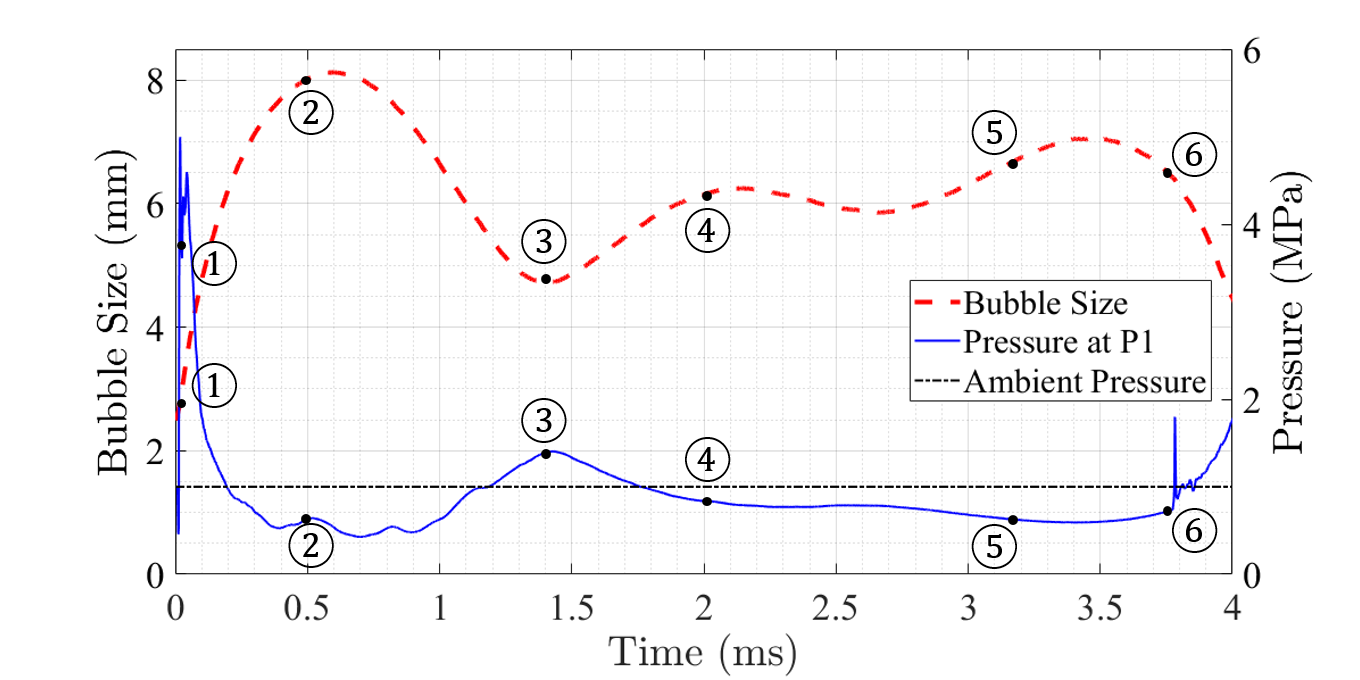}
 \caption{}
 \end{subfigure}
 \caption{Time history of selected quantities of interest in the case of $p_0 = 12.5~\text{MPa}$. (a) The distances between the top and bottom points and the left and right points of the cylinder. (b) The bubble size (radius equivalent) and the fluid pressure at a sensor location. The time instants shown in Figure~\ref{fig:12_5MPa_flowField} are marked on these curves.}
    \label{fig:12_5MPa_History}
\end{figure}

When the initial pressure inside the bubble is increased to $12.5$ MPa, the cylinder collapses in an orientation that features vertical extension and horizontal compression. This case has been briefly discussed in Section~\ref{subsec:meshConvergence} to demonstrate mesh convergence. Figure~\ref{fig:12_5MPa_flowField} shows six (6) solution snapshots, which illustrates the evolution of the bubble, the cylinder, and the fluid pressure and velocity fields. The cylinder's effective plastic strain is also visualized in the upper-row images. Again, the cylinder's horizontal and vertical widths are calculated to characterize its deformation (Figure~\ref{fig:12_5MPa_History}(a)). The time histories of bubble size and fluid pressure at a sensor location (P1 in Figure~\ref{fig:setup}) are shown in Figure~\ref{fig:12_5MPa_History}(b). 

In Figure~\ref{fig:12_5MPa_flowField}, the first snapshot is taken at $t = 0.022~\text{ms}$, the same time as Figure~\ref{fig:8MPa_flowField}\circled{1}. From the pressure field, it can be observed that the incident shock wave and its reflection both have a higher magnitude compared to the previous case ($p_0 = 8.0~\text{MPa}$). Again, the cylinder is compressed in the vertical direction due to both the shock load and the water flow generated by the expanding bubble. The compression stops at around $0.4$ ms, before the bubble reaches its maximum size (Figure~\ref{fig:12_5MPa_History}). Sub-figure \circled{2} is taken at $t = 0.493~\text{ms}$, when the structure is bouncing back in the vertical direction, while the bubble is still expanding. As a result, the downward expansion of the bubble is hindered by the structure. Unlike the previous case, a small amount of plastic deformation (effective plastic strain: $1.34\times 10^{-3}$) has already developed at this time at the bottom of the cylinder.  

The bubble reaches its maximum size at $t = 0.569$ ms. Then, it starts to contract. Figure~\ref{fig:12_5MPa_flowField}\circled{3} is taken at $t = 1.395~\text{ms}$, when the bubble reaches its minimum size. Again, a pressure pulse is generated by the bubble, which elevates the pressure around the cylinder (also see Figure~\ref{fig:12_5MPa_History}(b)). At this time, the cylinder is still stretched in the vertical direction and compressed in the horizontal direction (Figure~\ref{fig:12_5MPa_History}(a)). The elevated pressure field enhances its horizontal compression. Afterwards, the cylinder continues to deform in the same mode, instead of bouncing back as in the previous case. Sub-figures~\circled{3}\circled{4}\circled{5} illustrate this process. Therefore, the result suggests that the second pressure pulse generated by the contraction of the bubble have a significant impact on the cylinder's mode of collapse. 

Besides changing the collapse behavior of the cylinder, the increased initial pressure also influences the bubble dynamics through a complex dynamic interaction between the bubble and the cylinder. Like in the previous case, a dent forms at the bubble’s bottom surface. In the present case, this dent gradually evolves into a liquid “jet” during the process of the cylinder’s collapse. The jet penetrates the upper surface of the bubble at $3.46~\text{ms}$ (see Figure~\ref{fig:12_5MPa_flowField}, Sub-figures~\circled{5} and~\circled{6}). In the literature of cavitation, it is well-known that a bubble collapsing near a {\it rigid} surface often generates a liquid jet towards the surface, which can be an important mechanism of material damage~\cite{Turangan2017, Wang2017, Liu2017, Brujan2018, Cao2021}. It should be noted that the jet observed in the current simulation is in the opposite direction, and its formation is closely related to the deformation of the cylinder. This type of ``counter jet'' has also been observed previously in experiments that involve underwater explosion near an elastic solid body~\cite{Li2018}.

Figure~\ref{fig:12_5MPa_History}(a) shows that starting at around $1.2~\text{ms}$, the speed of the cylinder's horizontal compression keeps increasing. The collapse of the cylinder pulls the surrounding water towards it, which can be observed in the fluid velocity field in Figure~\ref{fig:12_5MPa_flowField}, Sub-figures~\circled{5} and~\circled{6}. At $t = 3.763~\text{ms}$, the cylinder reaches self-contact. An implosion shock wave is emitted at the point of contact because the inward motion of the surrounding water is suddenly stopped. This shock wave is also captured at sensor P1 (Figure~~\ref{fig:12_5MPa_History}(b)).

A comparison between Figures~\ref{fig:12_5MPa_History}(b) and~\ref{fig:8MPa_History}(b) reveals that as $p_0$ increases, the bubble's period of oscillation also increases. For example, the time when the bubble reaches the second maximum size is approximately $2.15~\text{ms}$ in the current case, compared to $1.76~\text{ms}$ in the previous case. This trend is consistent with simplified bubble dynamics models that assume spherical symmetry (e.g.~\cite{cole1948underwater, Brennen2013}).

\subsection{$p_0 = 25.0$~MPa (enthalpy $1.7181~\text{J/mm}$)}
\label{sec:25MPa}
\begin{figure}[!htb]
\hspace*{-2.5cm}  
 \includegraphics[width=170mm,trim={0cm 0cm 0cm 0cm},clip]{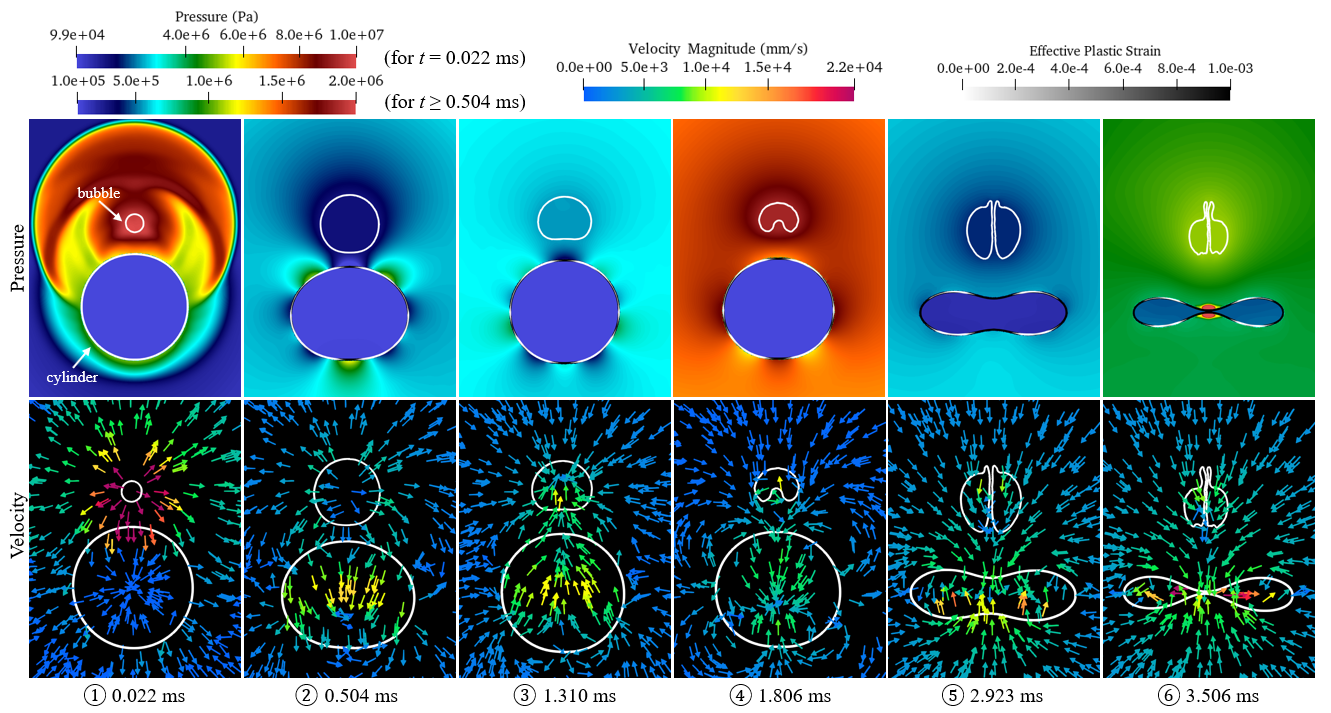}
    \caption{Snapshots of the fluid and structural results in the case of $p_0 = 25.0~\text{MPa}$.}
    \label{fig:25MPa_flowField}
\end{figure}

\begin{figure}[!htb]
\begin{subfigure}{.95\textwidth}
\centering
 \includegraphics[width=85mm,trim={0cm 0cm 0cm 0cm},clip]{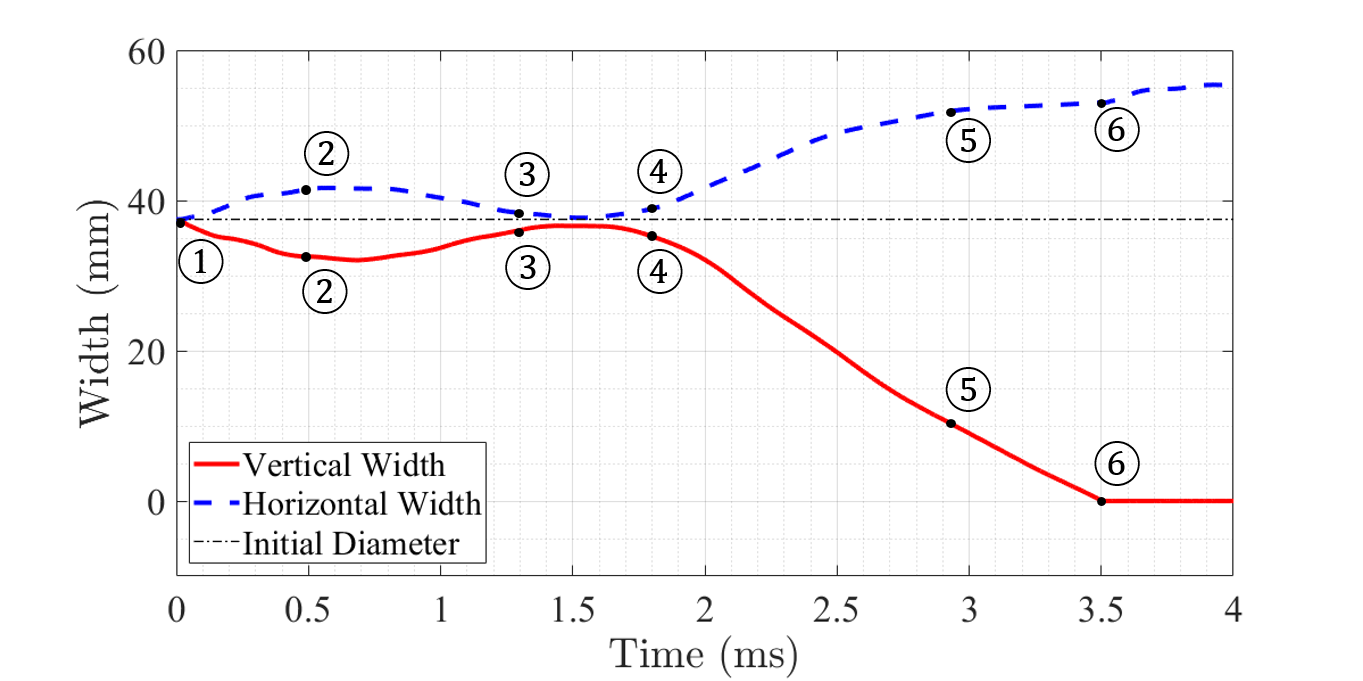}
 \caption{}
\end{subfigure}
\begin{subfigure}{.95\textwidth}
\centering
 \includegraphics[width=85mm,trim={0cm 0cm 0cm 0cm},clip]{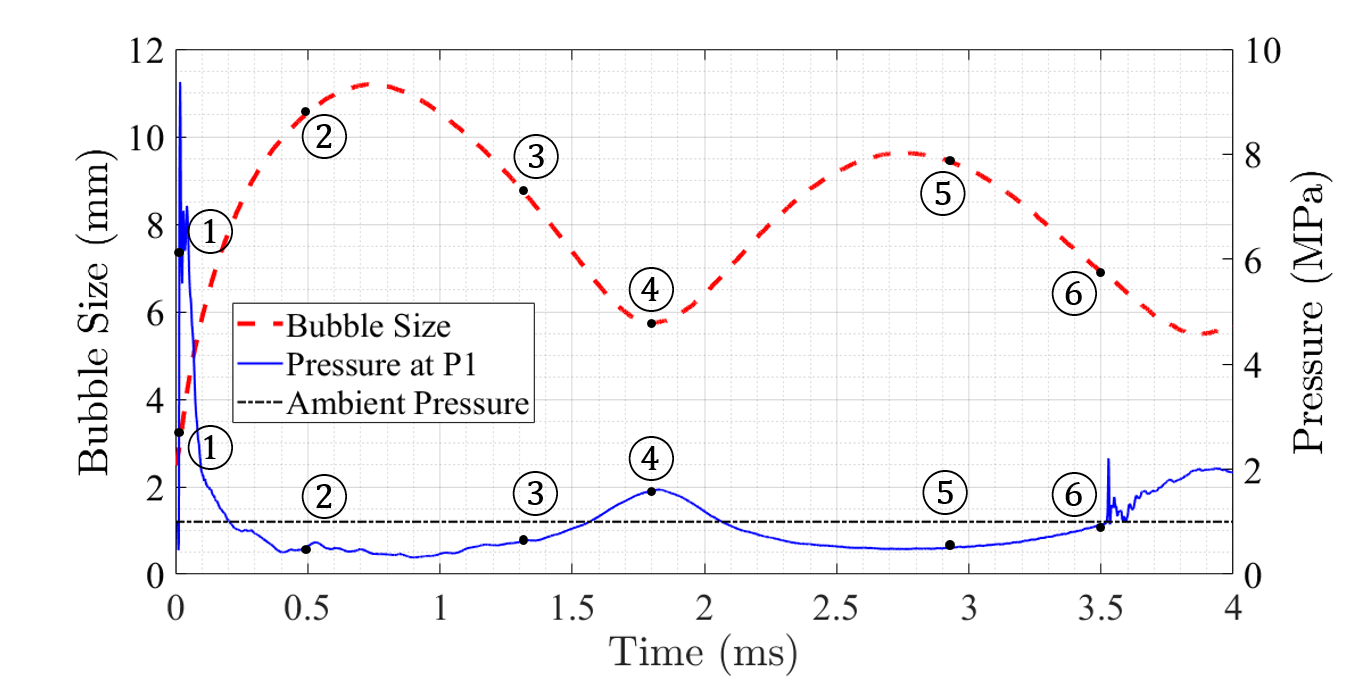}
 \caption{}
 \end{subfigure}
 \caption{Time history of selected quantities of interest in the case of $p_0 = 25.0~\text{MPa}$. (a) The distances between the top and bottom points and the left and right points of the cylinder. (b) The bubble size (radius equivalent) and the fluid pressure at a sensor location. The time instants shown in Figure~\ref{fig:25MPa_flowField} are marked on these curves.}
    \label{fig:25MPa_History}
\end{figure}

The initial pressure inside the bubble is increased further to $25.0~\text{MPa}$ in this case. Figure~\ref{fig:25MPa_flowField} presents six (6) solution snapshots. Figure~\ref{fig:25MPa_History} shows the time histories of the cylinder's deformation, the bubble size, and the fluid pressure at the same sensor location. It is found that the cylinder collapses in an orientation that is perpendicular to the one observed previously in the case of $12.5~\text{MPa}$. 

Due to the higher initial pressure inside the bubble, both the incident and the reflected shock waves have a higher magnitude (Figure~\ref{fig:25MPa_flowField}\circled{1}). Sub-figure~\circled{2} is taken at $t = 0.504~\text{ms}$, when the bubble is expanding and the cylinder being compressed in the vertical direction. At this time, plastic deformation has already developed at the top, bottom, left, and right points of the cylinder, which can be observed from the visualization of effective plastic strain. The cylinder's vertical compression stops at $t = 0.68~\text{ms}$, before the bubble reaches its maximum size. Then, the cylinder starts to bounce back. Unlike all the previous cases, in this case the cylinder cannot recover its original circular configuration, because of the developed plastic deformation. This is evident in Figure~\ref{fig:25MPa_History}(a), as the two curves never cross after $t = 0~\text{ms}$. Figure~\ref{fig:25MPa_flowField}\circled{3} is taken at $t = 1.310~\text{ms}$, when the cylinder is expanding vertically while the bubble is contracting. It can be seen that the bottom of the bubble is flattened by the flow induced by the cylinder's vertical expansion. The vertical expansion and horizontal contraction of the cylinder stop at $1.445~\text{ms}$. Then, the cylinder starts to deform in the opposite way.

Because of the increased initial pressure ($p_0$), the bubble's period of oscillation increases. It is at $t = 1.806~\text{ms}$ that the bubble contracts to its minimum size, compared to $t = 1.395~\text{ms}$ in the previous case. Sub-figure~\circled{4} in Figure~\ref{fig:25MPa_flowField} is taken at this time. Same as the previous cases, the contraction of the bubble generates a pressure pulse that elevates the pressure field around the cylinder. Nonetheless, in this case the delayed pressure pulse meets a cylinder that has a different configuration, that is, vertically compressed and horizontally stretched. As a result, the pressure pulse promotes the vertical compression of the cylinder. After this time, the cylinder loses stability and starts to collapse. Sub-figures~\circled{5} and~\circled{6} are taken during this process.

In this case, the bubble also produces a counter jet pointing away from the cylinder. This liquid jet keeps growing as the cylinder collapses, and it penetrates the upper surface of the bubble at $t = 2.509~\text{ms}$. Compared to the previous case, the jet is narrower and longer. When the cylinder is compressed vertically during the bubble's expansion phase after $1.806~\text{ms}$, the bubble's lower surface expands faster than other regions, which elongates the bubble downwards. Sub-figure~\circled{6} in Figure~\ref{fig:25MPa_flowField} is taken at the instant that the cylinder reaches self-contact. Again, the emission of an implosion shock wave can be clearly observed from the pressure field. It is notable that the collapsed configuration of the cylinder is symmetric with respect to its horizontal mid-plane (i.e.~the middle $x$-$z$ plane), despite that the external load is highly asymmetric. 

In addition, the cylinder collapses into a configuration that features vertical compression and horizontal expansion. Although this is a different mode compared to the previous case of $p_0 = 12.5~\text{MPa}$, the result shows that in both cases, the mode of collapse is determined at the time the bubble reaches its minimum size. The difference in collapse mode can be explained by the different configurations of the cylinder at this time.
Specifically, in this case, the cylinder is vertically compressed and horizontally stretched, whereas in the case with $p_0 = 12.5~\text{MPa}$, it is vertically stretched and horizontally compressed. Furthermore, the result suggests that the difference in the cylinder's configuration at the arrival of the pressure pulse is related to both the increase of the bubble's period of oscillation and the cylinder's plastic deformation.

\subsection{$p_0 = 50.0$~MPa (enthalpy $3.4361~\text{J/mm}$)}
\label{sec:50MPa}

\begin{figure}[!htb]
\hspace*{-2.5cm}  
 \includegraphics[width=170mm,trim={0cm 0cm 0cm 0cm},clip]{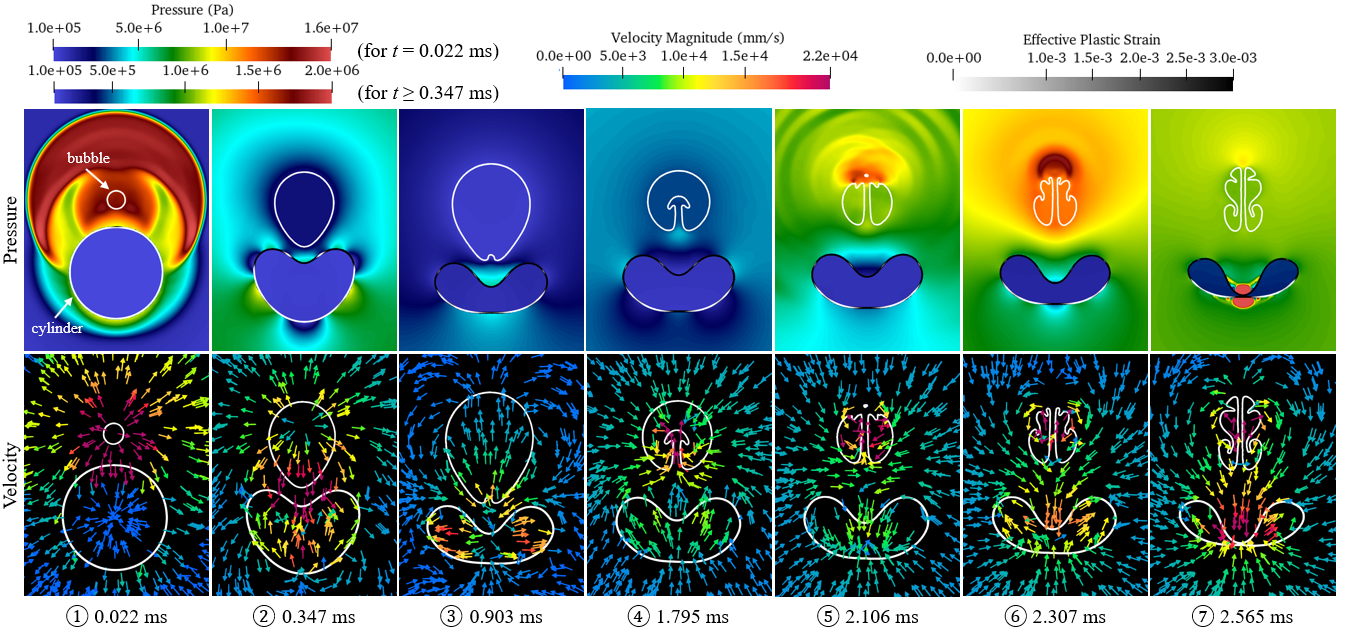}
    \caption{Snapshots of the fluid and structural results in the case of $p_0 = 50.0~\text{MPa}$.}
    \label{fig:50MPa_flowField}
\end{figure}

\begin{figure}[!htb]
\begin{subfigure}{.95\textwidth}
\centering
 \includegraphics[width=85mm,trim={0cm 0cm 0cm 0cm},clip]{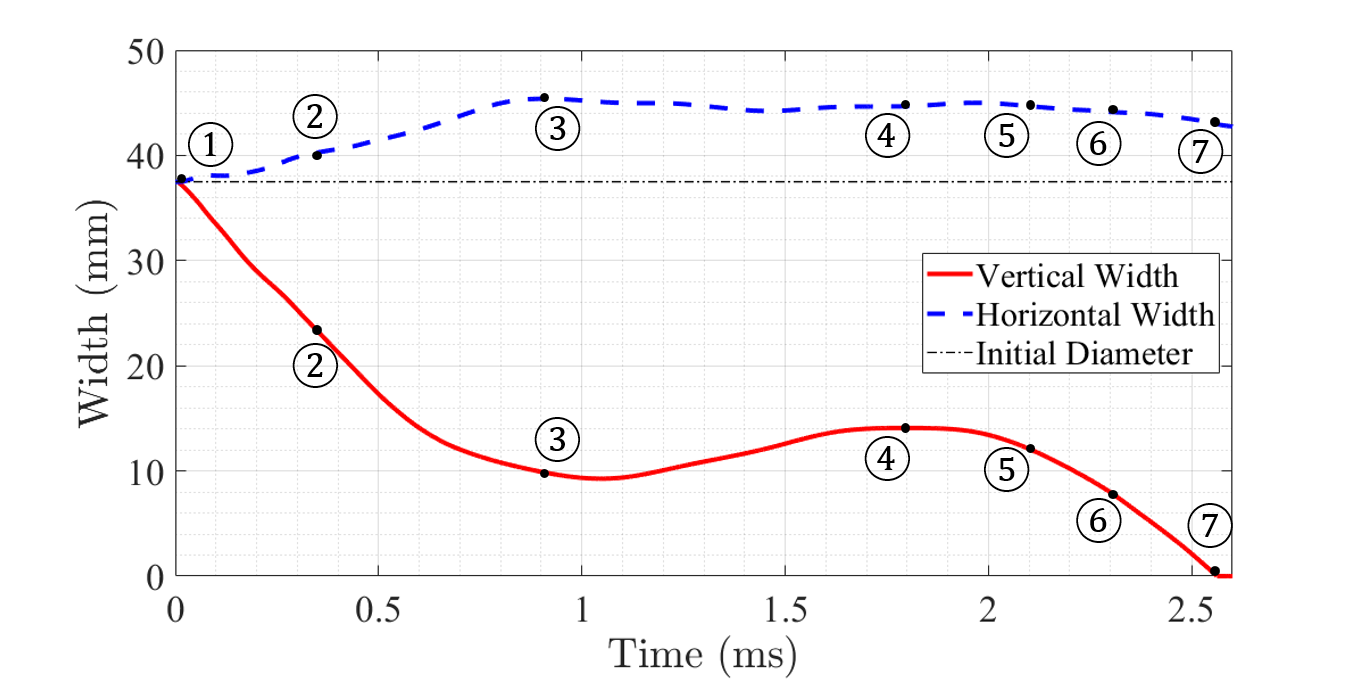}
 \caption{}
\end{subfigure}
\begin{subfigure}{.95\textwidth}
\centering
 \includegraphics[width=85mm,trim={0cm 0cm 0cm 0cm},clip]{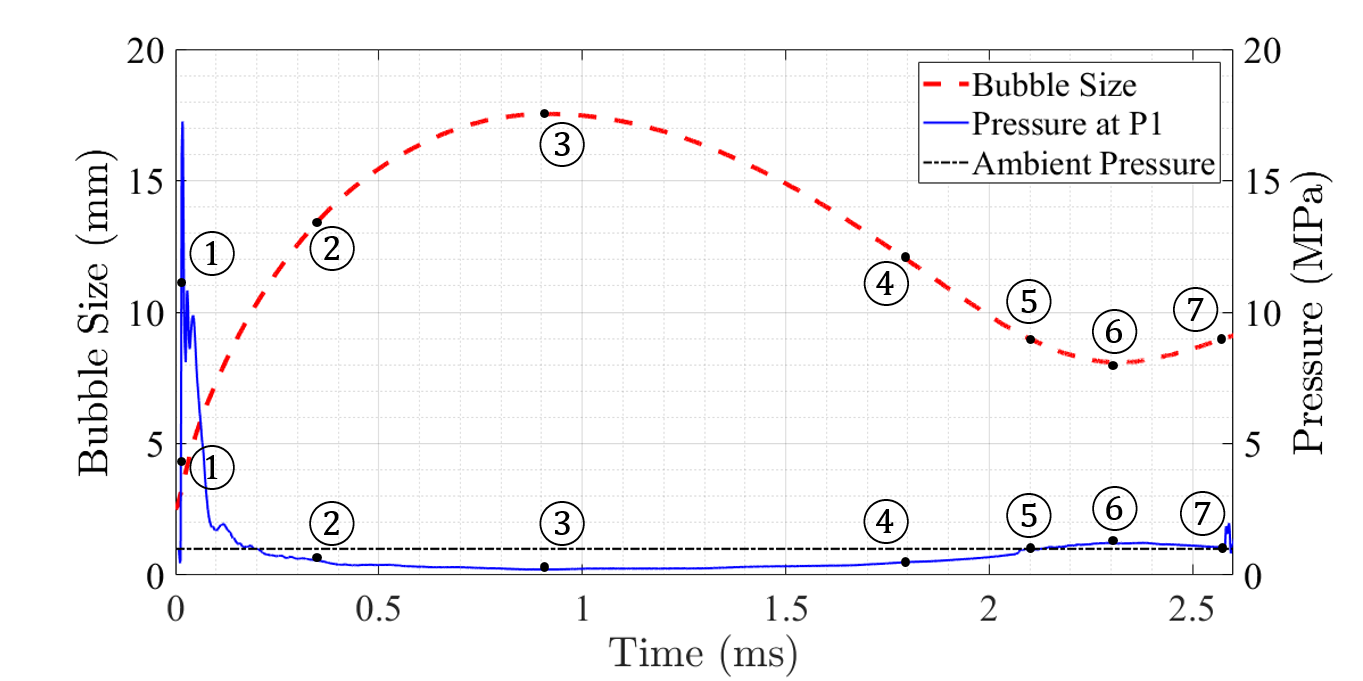}
 \caption{}
 \end{subfigure}
 \caption{Time history of selected quantities of interest in the case of $p_0 = 50.0~\text{MPa}$. (a) The distances between the top and bottom points and the left and right points of the cylinder. (b) The bubble size (radius equivalent) and the fluid pressure at a sensor location. The time instants shown in Figure~\ref{fig:50MPa_flowField} are marked on these curves.}
    \label{fig:50MPa_History}
\end{figure}

In this case, the cylinder collapses into another shape that is noticeably different from those observed in the cases of $p_0 = 12.5~\text{MPa}$ and $p_0 = 25~\text{MPa}$. The same set of results are extracted and presented in Figures~\ref{fig:50MPa_flowField} and~\ref{fig:50MPa_History}.

In Figure~\ref{fig:50MPa_flowField}, sub-figure~\circled{1} is taken at the same time as in the previous cases, which highlights the initial loads from the bubble including a shock wave with pressure of the order of $20~\text{MPa}$ and a water flow with velocity of the order of $4\times10^4~\text{mm}/\text{s}$. At $t = 0.347~\text{ms}$ (Sub-figure~\circled{2}), the result is already very different from the previous cases. Both the bubble and the cylinder are no longer symmetric with respect to their horizontal mid-planes (i.e.~the middle $x$-$z$ plane). The top region of the cylinder is collapsing, while the bottom region has much smaller, elastic deformation. Despite a higher initial pressure ($p_0$) compared to the previous case, yielding in the bottom region of the cylinder is delayed. Sub-figure~\circled{2} also shows that the bubble deforms into an oval shape, as its bottom region is pulled by a downward velocity field that is in accordance with the collapse of the cylinder. 

Sub-figure~\circled{3} is taken at $t = 0.903~\text{ms}$. Around this time, the collapse of the cylinder stops, despite that its vertical width has dropped by over $70\%$ compared to the original configuration (Figure~\ref{fig:50MPa_History}(a)). Figure~\ref{fig:50MPa_History}(b) shows that this is the time when the bubble has reached its maximum size and starts to contract. The contraction of the bubble is an inertial effect, caused by the continuous decrease of the internal pressure during the bubble's expansion. It pulls the surrounding water towards it, thereby facilitating the formation of the velocity field that stops the downward collapse of the cylinder.  Similar to the previous cases, a dent can be observed at the bottom of the bubble, which gradually evolves into an upward liquid jet. Sub-figure~\circled{4} is taken at $t = 1.795~\text{ms}$, when the collapse of the cylinder resumes. As the cylinder collapses, the liquid jet continues penetrating the bubble. Sub-figure~\circled{5} is taken at $t = 2.106~\text{ms}$, shortly after the jet penetrates the top surface of the bubble. At $t = 2.307~\text{ms}$, the bubble contracts to its minimum size (Sub-figure~\circled{6}). Finally, at $t = 2.565~\text{ms}$, the cylinder reaches self-contact (Sub-figure~\circled{7}). Again, an implosion shock wave is generated at the point of contact.

In summary, because of the higher initial pressure ($p_0$), the collapse of the cylinder starts at an earlier time compared to the previous two cases ($12.5~\text{MPa}$ and $25~\text{MPa}$), and it starts only in the top region, which is close to the bubble. In addition, the collapse of the cylinder does not progress in a monotonic fashion. Instead, it is temporarily pulled back as the bubble contracts, which again indicates a strong coupling between the bubble dynamics and the cylinder's transient deformation. At the end, when the cylinder reaches self-contact, its configuration is clearly different from the shape observed in the case of $p_0 = 25~\text{MPa}$ in that it is no longer symmetric with respect to the horizontal mid-plane (i.e.~the middle $x$-$z$ plane).


\subsection{$p_0 = 100.0$~MPa (enthalpy $6.8722~\text{J/mm}$)}
\label{sec:100MPa}
\begin{figure}[!htb]
\hspace*{-2.5cm}  
 \includegraphics[width=170mm,trim={0cm 0cm 0cm 0cm},clip]{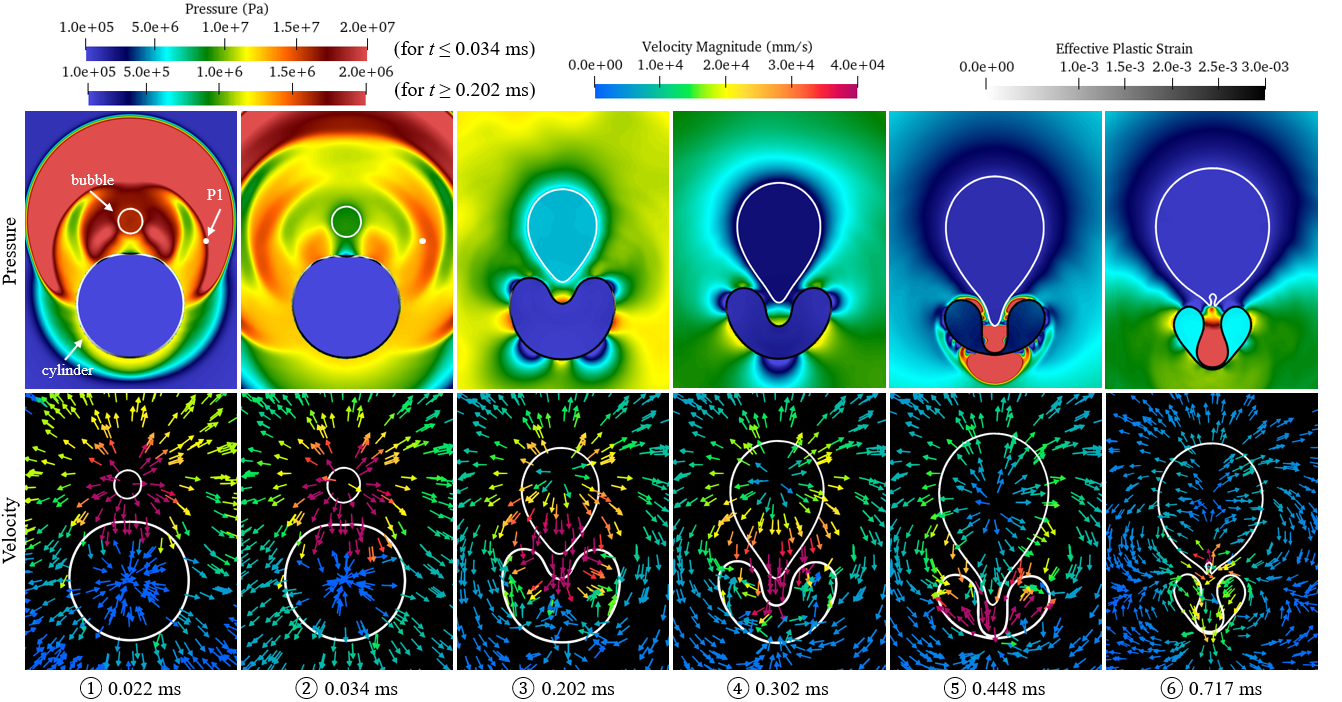}
    \caption{Snapshots of the fluid and structural results in the case of $p_0 = 100.0~\text{MPa}$. The sensor location P1 is marked in the first two sub-figures.}
    \label{fig:100MPa_flowField}
\end{figure}

\begin{figure}[!htb]
\begin{subfigure}{.95\textwidth}
\centering
 \includegraphics[width=85mm,trim={0cm 0cm 0cm 0cm},clip]{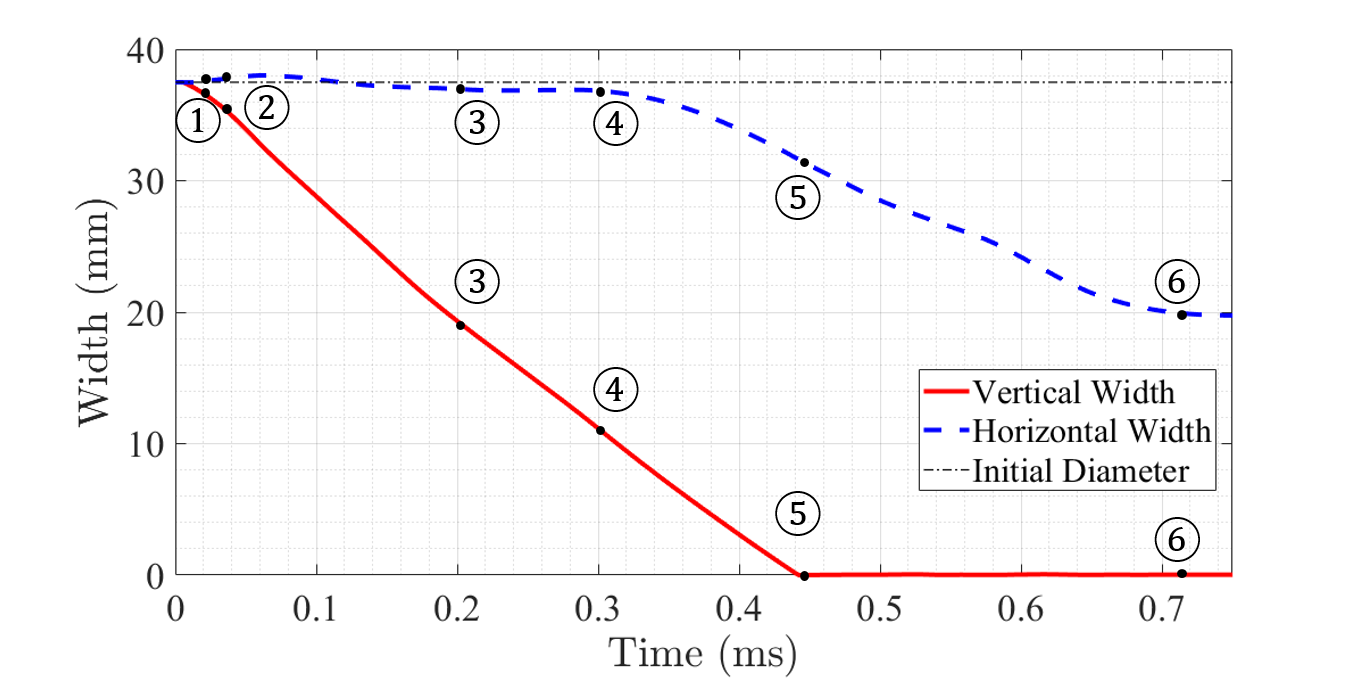}
 \caption{}
\end{subfigure}
\begin{subfigure}{.95\textwidth}
\centering
 \includegraphics[width=85mm,trim={0cm 0cm 0cm 0cm},clip]{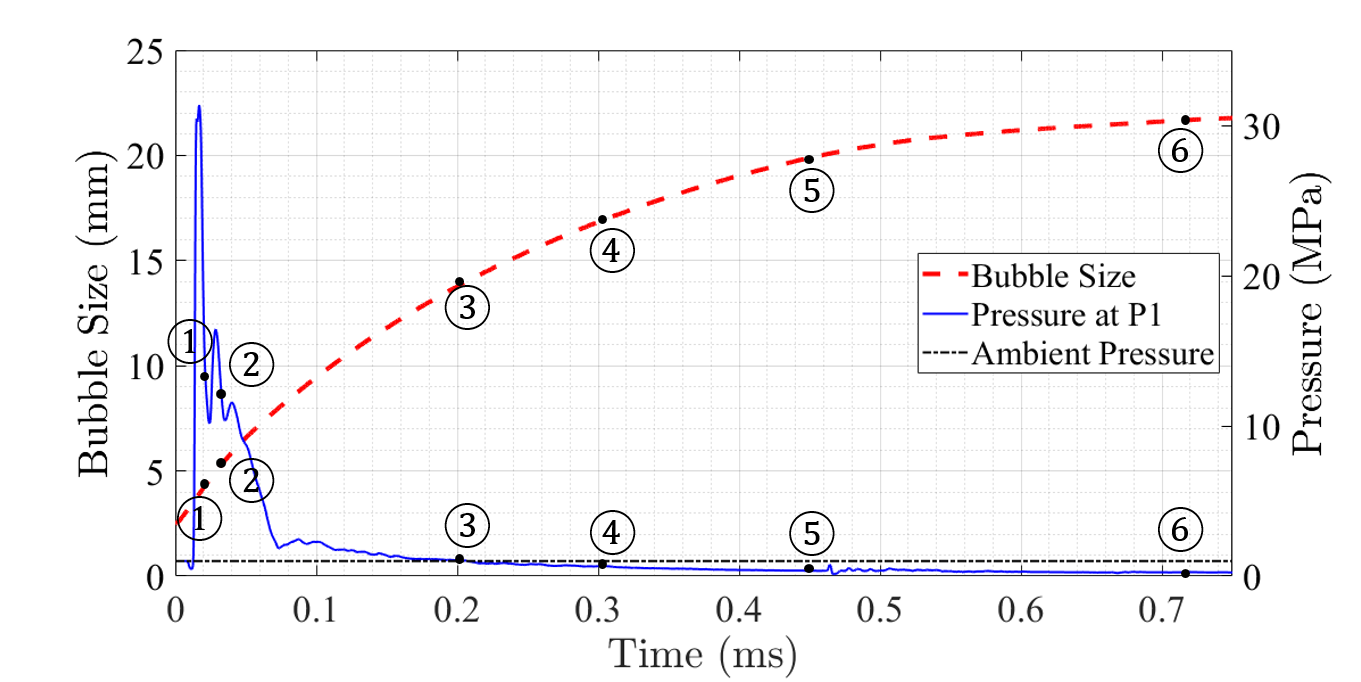}
 \caption{}
 \end{subfigure}
 \caption{Time history of selected quantities of interest in the case of $p_0 = 100.0~\text{MPa}$. (a) The distances between the top and bottom points and the left and right points of the cylinder. (b) The bubble size (radius equivalent) and the fluid pressure at a sensor location. The time instants shown in Figure~\ref{fig:100MPa_flowField} are marked on these curves.}
    \label{fig:100MPa_History}
\end{figure}

In this case, the cylinder collapses into a configuration similar to the one observed in the previous case. The main difference is that the collapse progresses monotonically, and is not interrupted by the bubble. The results are presented in Figures~\ref{fig:100MPa_flowField} and~\ref{fig:100MPa_History}.

In Figure~\ref{fig:100MPa_flowField}, Sub-figure~\circled{1} is taken at 0.022 ms, the same time instant as in the previous cases. At this time, yielding has started not only in the top region of the cylinder, but also in its bottom region. The pressure field shows that the incident shock wave (i.e.~the region that has red color and a crescent shape) has not reached the bottom of cylinder through water. Therefore, the yielding of the bottom region should be attributed to the convergence of the stress waves propagating downward along the two sides of the cylinder. Sub-figure~\circled{2} is taken at $t = 0.034~\text{ms}$, when the reflected shock wave has just passed the sensor location P1. This reflection is captured by the sensor as the second, smaller pressure spike (Figure~\ref{fig:100MPa_History}(b)). At this time, plastic deformation has already developed around the entire circumference of the cylinder. Sub-figure~\circled{3} is taken at $0.202~\text{ms}$. At this time, the top region of the cylinder is highly concave. Again, the bottom region of the bubble is pulled by the downward velocity field created by the collapsing cylinder. As a result, it deforms into an oval shape. Sub-figure~\circled{4} is taken at $0.302~\text{ms}$. From Figure~\ref{fig:100MPa_History}(a) it can be observed that until this time, the horizontal width of the cylinder has largely remained constant. Afterwards, it starts to decrease, as the two lobes of the cylinder fold towards each other. This is different from the last two cases ($p_0 = 25~\text{MPa}$ and $50~\text{MPa}$) as in those cases, the horizontal width increases as the cylinder collapses. At $t = 0.448~\text{ms}$ (sub-figure~\circled{5}), the cylinder reaches self-contact and emits an implosion shock wave. The bubble is still expanding at this time instant. Lastly, Sub-figure~\circled{6} is taken at $t = 0.717~\text{ms}$. Again, a liquid jet forms at the bottom of the bubble, and it will gradually penetrate the bubble.

In this case, the cylinder collapses within $0.5$ ms, a time interval that is much shorter than the previous cases. On the other hand, the bubble's period of oscillation is longer because of the higher initial pressure ($p_0$). As a result of these changes, the bubble keeps expanding during the entire collapsing process of the cylinder (Figure~\ref{fig:100MPa_History}(b)). Since bubble contraction does not happen during the collapse of the cylinder, the cylinder does not rebound like in the previous case ($p_0 = 50~\text{MPa}$).    

\subsection{Summary}

\begin{figure}[!htb]
    \centering
\hspace*{-1cm}  
 \includegraphics[width=140mm,trim={0cm 0cm 0cm 0cm},clip]{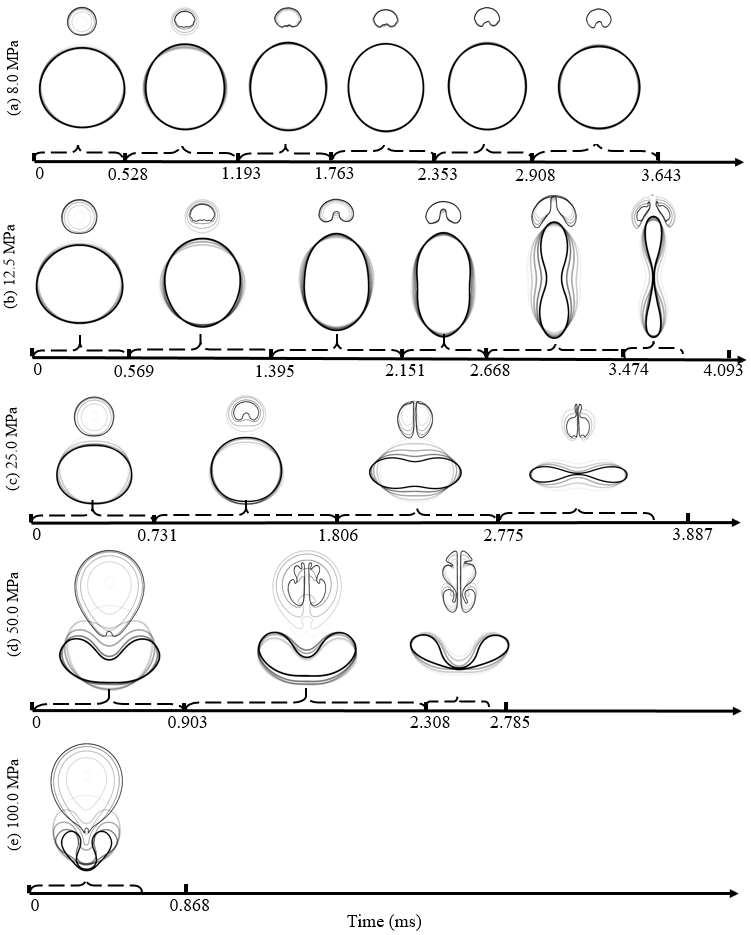}
    \caption{Comparison of the evolution of the bubble and the cylinder in the five ($5$) representative cases. One image is generated for each half cycle (expansion or contraction) of the bubble. Within each image, results at different time instants are superimposed, and the increase of opacity indicates the time evolution.}
    \label{fig:summary}
\end{figure}

We have discussed five ($5$) representative cases with different initial pressure inside the bubble ($p_0$). The results show that the collapse behavior of the cylinder can be drastically different as $p_0$ varies, while all the other parameters remain fixed. Figure~\ref{fig:summary} presents a comparison of all the five cases in terms of structural deformation and bubble dynamics. In this figure, the time axis is synchronized among all the cases. For each case, the ticks on the time axis mark the start and end times of the bubble's half cycles,~i.e. the expansion and contraction phases. One image is presented within each time interval, in which the bubble and the cylinder's configuration at different time instants are superimposed. The time evolution is shown using opacity. Specifically, a darker line corresponds to a result at a later time. The time interval between adjacent time instants is fixed within each image. The dashed curly brackets along the time axis represent the time span of each image. 

In the first case ($p_0 = 8.0~\text{MPa}$), due to the low pressure inside the bubble, the cylinder vibrates without collapsing. As $p_0$ increases, in the second case ($p_0 = 12.5~\text{MPa}$) the cylinder collapses into a configuration that features horizontal compression and vertical extension. The top of the cylinder, which is closest to the bubble, is found to move towards the bubble. In the third case ($p_0 = 25.0~\text{MPa}$), the cylinder collapses into a configuration that features vertical compression and horizontal extension, that is, a configuration perpendicular to that observed in the second case. Notably, in both cases, the collapsed configuration is symmetric with respect to the cylinder's horizontal mid-plane (i.e.~the middle $x$-$z$ plane), despite the fact that the loading is clearly asymmetric. In the fourth case ($p_0 = 50.0~\text{MPa}$), the cylinder is still vertically compressed after collapsing but the aforementioned symmetry is lost. The deformation mostly occurs at the top region of the cylinder. Moreover, the cylinder does not collapse monotonically. It is pulled back by the bubble during a short period of time when bubble is contracting. In the last case ($p_0 = 100.0~\text{MPa}$), the cylinder collapses monotonically within a very short period of time. It reaches self-contact before the bubble completes the first expansion phase.  

\section{Transition of collapse modes}
\label{sec:transition}

In this section, we discuss the different types of collapse behaviors using a bigger data set that consists of $15$ simulations in which $p_0$ is varied from $1.0~\text{MPa}$ to $100.0~\text{MPa}$. The transition among different modes of collapse are investigated. 

\subsection{Different collapse behaviors}
\label{sec:differentCollapse}
%
%
%
%
%
%

\begin{figure*}[!htb]
 \includegraphics[width=120mm,trim={0cm 0cm 0cm 0cm},clip]{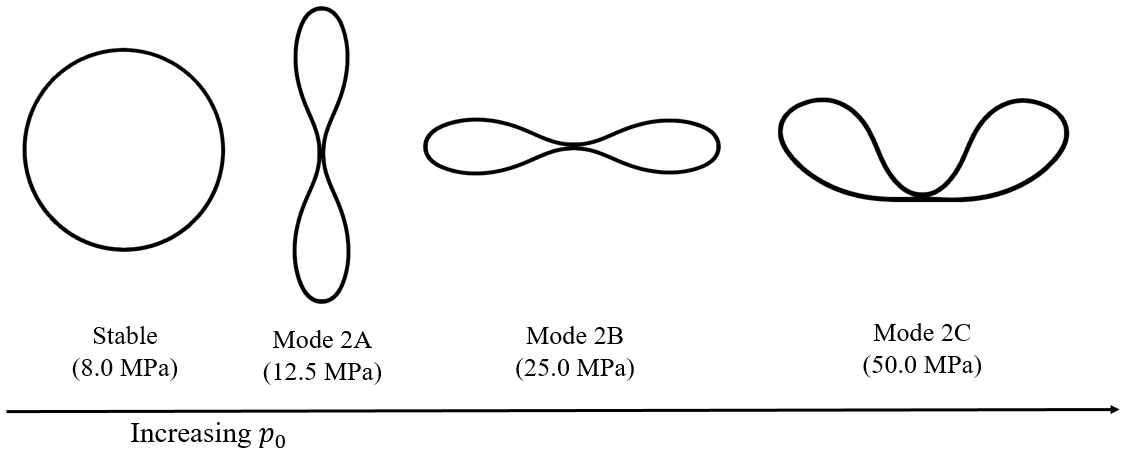}
    \caption{The transition of collapse modes as $p_0$ increases from $1~\text{MPa}$ to $100~\text{MPa}$.}
    \label{fig:modes}
\end{figure*}

\begin{figure*}[!htb]
\centering
\hspace*{-2.5cm} 
\includegraphics[width=170mm,trim={0cm 0cm 0cm 0cm},clip]{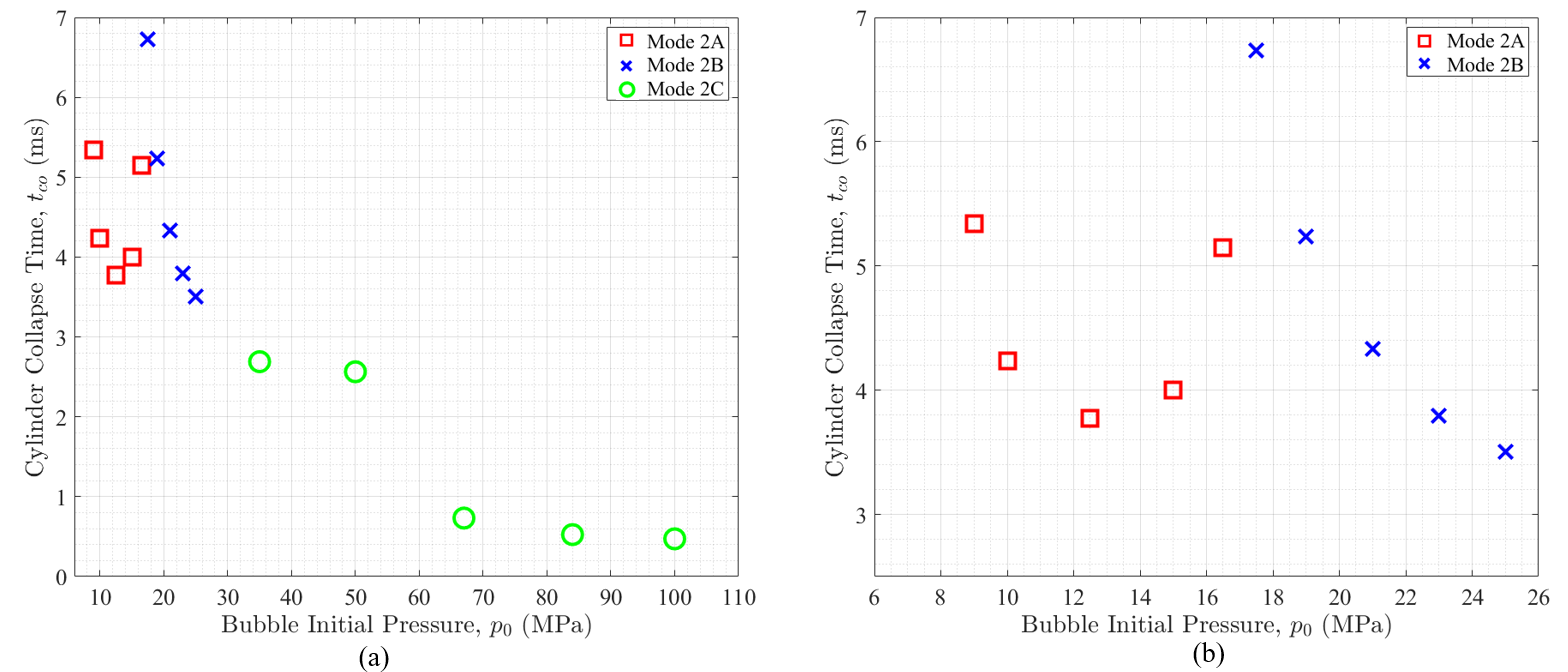}
\caption{The time to collapse ($t_{co}$) for (a) $8~\text{MPa}~<~p_0 \leq 100.0~\text{MPa}$ and (b) $8~\text{MPa}~<~p_0 \leq 25~\text{MPa}$.}
\label{fig:collapseTime}
\end{figure*}

Figure~\ref{fig:modes} summarizes the different collapse behaviors observed in this parametric study. In the remainder of this paper, the mode of collapse that features symmetric horizontal compression is referred to as Mode 2A. The mode of collapse that features symmetric vertical compression is referred to as Mode 2B. The asymmetric collapse with vertical compression is referred to as Mode 2C. Here, the number $2$ refers to the fact that the collapsed configuration contains two lobes. In addition, Figure~\ref{fig:collapseTime} presents the mode of collapse observed in each test case, as well as the time it takes for the cylinder to reach self-contact, denoted by $t_{co}$.

A few findings from this parametric study are noteworthy.

\begin{enumerate}
\item[(1)] As pressure $p_0$ increases, the first collapse mode (Mode 2A) features horizontal compression and vertical extension.
\item[(2)] The time to collapse, $t_{co}$, is not a monotonically decreasing function of $p_0$, despite that a larger $p_0$ generally means a stronger load.
\item[(3)] As $p_0$ increases beyond $16.5$ MPa, a change of collapse mode (from Mode 2A to Mode 2B) is observed.
\end{enumerate}


\subsection{Discussion}

\begin{figure} [!htb]
    \centering
    \hspace*{-2.2cm} 
    \includegraphics[width=170mm, trim={0cm 0cm 0cm 0cm}, clip]{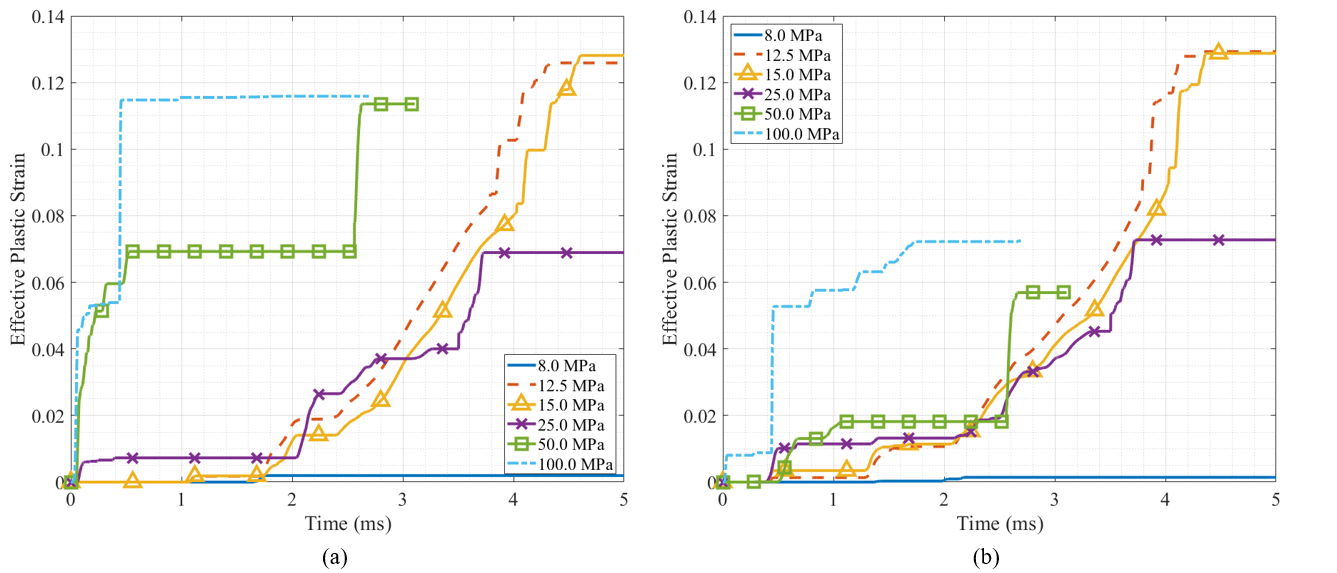}
    \caption{Effective plastic strain at the (a) top point and (b) bottom point of the cylinder's inside wall.}
    \label{fig:EPS_topBottom}
\end{figure}

To explain the findings mentioned above, it is helpful to examine the time history of plastic strain in the cylinder. Figure~\ref{fig:EPS_topBottom} presents the effective plastic strain measured at the top and bottom points of the cylinder's inside wall. From Figure~\ref{fig:EPS_topBottom}(a), it is clear that Mode 2A collapse is not induced directly by the initial loads,~i.e.~the incident shock wave and the water flow caused by the bubble's initial expansion. For example, in the cases of $p_0 = 12.5~\text{MPa}$ and $15.0~\text{MPa}$, the effective plastic strain remains zero (or nearly zero) until $1~\text{ms}$, when the initial loads have long passed. In both cases, plastic strain starts to develop after $1.5~\text{ms}$, when the surrounding water pressure is elevated due to the bubble's contraction (cf.~Figures~\ref{fig:12_5MPa_flowField} and~\ref{fig:12_5MPa_History}). Therefore, the first collapse mode is induced mainly by the first contraction phase of the bubble and the resulting pressure pulse. Around this time, the cylinder happens to be in a configuration of vertical extension and horizontal compression, which determines the shape of Mode 2A collapse.

\begin{figure} [!htb]
    \centering
    \hspace*{-2.2cm} 
    \includegraphics[width=170mm, trim={0cm 0cm 0cm 0cm}, clip]{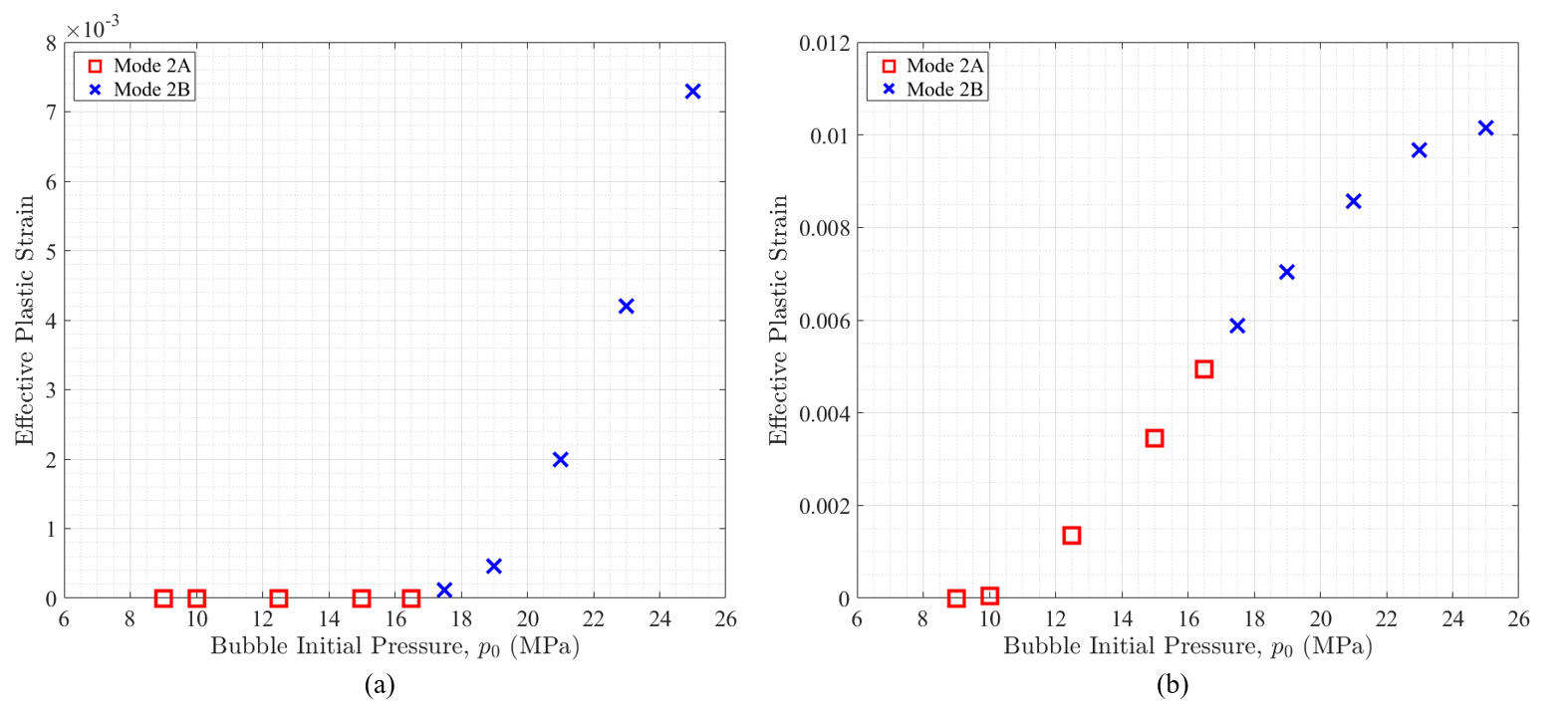}
    \caption{Effective plastic strain at the cylinder inside wall's (a) top and (b) bottom points when the cylinder's vertical width reaches minimum for the first time (i.e.~when the cylinder completes the first half-cycle of vibration).}
    \label{fig:epsEarly}
\end{figure}

Figure~\ref{fig:collapseTime}(b) shows that for the cases that result in Mode 2A collapse, the time to collapse ($t_{co}$) does not decrease monotonically as $p_0$ increases. Specifically, when $p_0$ is increased from $12.5~\text{MPa}$ to $15.0~\text{MPa}$, the cylinder collapses slower. This phenomenon is related to the plastic deformation caused by the initial loads. Figure~\ref{fig:EPS_topBottom}(b) shows that, in the cases of $p_0 = 12.5~\text{MPa}$ and $p_0 = 15.0~\text{MPa}$, there is a small amount of plastic deformation at the bottom of the cylinder around $0.5~\text{ms}$. Around this time, the cylinder's vertical width reaches a minimum value. In other words, the cylinder has just completed the first half-cycle of vibration, which results directly from the initial loads (cf.~Figure~\ref{fig:12_5MPa_History}(a)). Figure~\ref{fig:epsEarly} provides a cross comparison of the plastic deformation induced by the initial loads among all the test cases. It is clear that as $p_0$ increases, the effective plastic strain also increases. As discussed in Section~\ref{sec:25MPa}, such plastic deformation tends to hinder Mode 2A collapse. Therefore, the result shows that the increased plastic deformation due to stronger initial loads is a factor that increases the cylinder's time to collapse ($t_{co}$) in Mode 2A. 

\begin{figure} [!htb]
    \centering
    \includegraphics[width=85mm, trim={0cm 0cm 0cm 0cm}, clip]{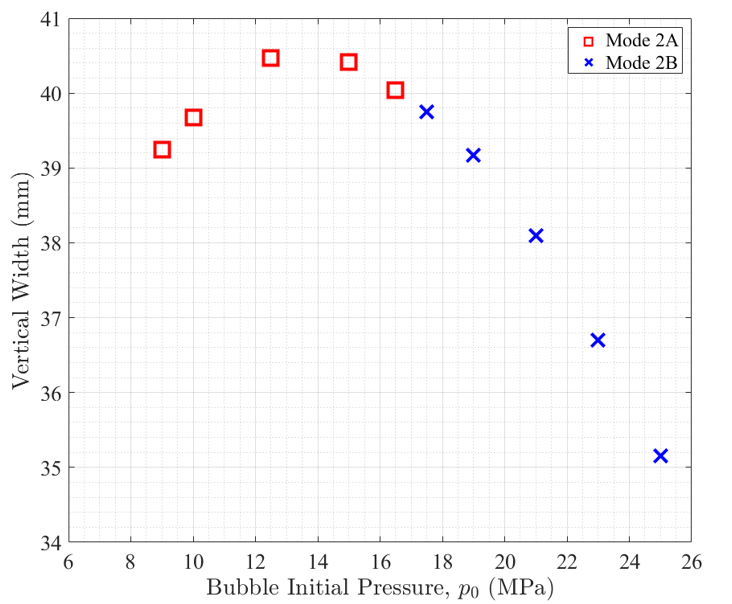}
    \caption{Vertical width of the cylinder when the bubble contracts to its minimum size.}
    \label{fig:verticalWidthAll}
\end{figure}

The increase of plastic deformation caused by the initial loads also leads to the fact that when the cylinder completes its first cycle of vibration, it can no longer recover the initial configuration. Instead, it is vertically compressed and horizontally stretched. Figure~\ref{fig:verticalWidthAll} shows that as $p_0$ further increases beyond $12.5~\text{MPa}$, the vertical width of the cylinder decreases monotonically. As described in Section~\ref{sec:examples}, when the cylinder completes its first cycle of vibration, it is subjected to a pressure pulse that results from the contraction of the bubble. This pressure pulse drives the cylinder to collapse in the vertical direction. In summary, the results suggest that Mode 2B collapse is triggered by the initial loads and facilitated by the bubble's dynamics, particularly its first contraction phase that generates the second pressure pulse.

In addition of the findings listed in Section~\ref{sec:differentCollapse}, in Figure~\ref{fig:collapseTime}, it can be seen that when the cylinder collapses in Mode 2C, there is a sudden decrease of $t_{co}$ between $p_0 = 50.0~\text{MPa}$ and $67.0$~MPa. This behavior is related to the time sequence of cylinder collapse and bubble contraction, as discussed in Section~\ref{sec:50MPa} and Section~\ref{sec:100MPa}.

\section{Conclusion}
\label{sec:conclusion}

The collapse of an underwater aluminum cylinder due to a near-field explosion is investigated using fluid-structure coupled simulations. Previous studies in this area suggest that the dynamics of the explosion bubble may have a substantial effect in this type of events. Nonetheless, knowledge about this effect is very limited. Therefore, a specific objective of this study is to capture and explain the two-way interaction between the explosion bubble and the structure, using a two-dimensional model problem. 

The computational model employed in this study combines a multiphase compressible fluid dynamics solver with a nonlinear structural dynamics solver. It has been verified and validated for several problems that are closely related to the current work, including the collapse of aluminum cylinders due to hydrostatic pressure, and the pulsation of bubbles in free field and near solid materials. In this work, we start with a mesh convergence analysis in which the fluid and structural meshes are progressively refined until convergence is achieved. Afterwards, a parametric study is conducted, in which the initial pressure inside the explosion bubble ($p_0$) is varied by two orders of magnitude. The interaction of the bubble, the surrounding liquid water, and the aluminum cylinder is investigated by examining the fluid pressure and velocity fields, the bubble dynamics, and the transient structural deformation and stresses. 

It is found that as $p_0$ varies, the final configuration of the cylinder can be substantially different. Results from five representative cases ($p_0 = 8~\text{MPa}$, $12.5~\text{MPa}$, $25~\text{MPa}$, $50~\text{MPa}$, $100~\text{MPa}$) are discussed in detail. In these cases, the structural dynamics varies from a cyclic elastic vibration without collapse ($p_0 = 8~\text{MPa}$) to an immediate collapse without vibration ($p_0 = 100~\text{MPa}$). Three different types of collapse behaviors are observed, which are categorized as Mode 2A, Mode 2B, and Mode 2C (Figure~\ref{fig:modes}). As $p_0$ increases, the mode of collapse changes from 2A to 2B, and then from 2B to 2C. The mode transitions are discussed using additional test cases in the parametric study.

The mechanisms of the three collapse modes are summarized below. Mode 2A is caused by a coincidence between the bubble's first contraction phase and the cylinder's horizontal compression. As the bubble contracts, it emits a pressure pulse that elevates the pressure around the entire cylinder. When this bubble pulse arrives at the horizontally compressed cylinder, it facilitates its compression, leading to an horizontal collapse that has been observed earlier in a laboratory experiment~\cite{Ikeda2012}. Mode 2B is related to the plastic deformation induced by the initial loads, i.e., the initial shock wave and the water flow caused by the bubble's initial expansion. This plastic deformation suppresses the vertical extension and horizontal compression of the cylinder. As a result, when the aforementioned bubble pulse arrives at the cylinder, the cylinder is compressed in vertical direction. The bubble pulse facilitates this compression, leading to collapse mode 2B. For Mode 2C collapse, the result suggests that the cylinder's collapse is directly induced by the incident shock wave from the explosion. Clearly, the dynamics of the explosion bubble has a significant effect in collapse Modes 2A and 2B. In both cases, the plastic deformation remains relatively small long after the initial loads have passed. The collapse of the cylinder starts only after the bubble contracts to its first minimum size and emits a pressure pulse. 
 
The time that the cylinder takes to completely collapse does not decrease monotonically as $p_0$ increases (cf. Figure~\ref{fig:collapseTime}). When the cylinder collapses in Mode 2A, increasing $p_0$ may cause the cylinder to take more time to reach self-contact. This phenomenon is caused by the increased amount of plastic deformation induced by the initial loads from the explosion.
 
The simulation result also reveals that the dynamics of an explosion bubble near a vibrating or collapsing cylinder is significantly different from the dynamics of bubbles in free field or near a rigid wall. In other words, the transient structural deformation has a clear effect on the bubble dynamics. In particular, in the cases with $p_0 \geq 8~\text{MPa}$, a counter-jet that points away from the structural surface is observed. The formation of this counter-jet is induced by the vibration and collapse of the cylinder. This type of phenomenon has been observed previously in bubbles expanding near an elastic solid body. Compared to the liquid jets produced by bubbles collapsing near a rigid wall, this counter-jet is in the opposite direction.

\section*{Declaration of competing interest}

The authors have no relevant financial or non-financial competing interests.

\section*{Acknowledgement}

W.M., X.Z., and K.W. gratefully acknowledge the support  of the Office of Naval Research (ONR) under Awards N00014-19-1-2102, and the support of the National Science Foundation (NSF) under Awards CBET-1751487. C.G. gratefully acknowledges ONR N00014-10-C-0108 and Prof. James H. Duncan from the University of Maryland.

\end{document}